\documentclass{vldb_nocopyright}

\usepackage{booktabs} 
\usepackage[inline]{aplcomments} 
\usepackage{xspace} 
\usepackage{enumitem}
\usepackage{listings}
\usepackage{xcolor}
\usepackage{dsfont}
\usepackage{amsmath}
\usepackage{stmaryrd}
\usepackage{amsfonts}
\usepackage{mathtools}
\usepackage{algpseudocode}
\usepackage{algorithm}
\usepackage{inconsolata}
\usepackage{esvect}
\usepackage{hyperref}
\usepackage{adjustbox}
\usepackage{url}
\makeatletter
\g@addto@macro{\UrlBreaks}{\UrlOrds}
\makeatother

\newcommenter{chu}{1.0,0.4,1.0} 
\newcommenter{dan}{1.0,0.4,0.4} 
\newcommenter{alvin}{0.4,1.0,1.0} 

\definecolor{forestgreen}{RGB}{64,139,64}
\lstdefinestyle{sql}{ 
language=sql,
basicstyle=\scriptsize\ttfamily, 
breakatwhitespace=false, 
breaklines=true, 
captionpos=b, 
commentstyle=\itshape\color[rgb]{0.4,0.4,0.4}, 
deletekeywords={VALUE, INPUT, SIZE}, 
firstnumber=1, 
frame=none, 
frameround=tttt, 
keywordstyle=\color[RGB]{0, 110, 184},
morekeywords={SEMIJOIN, WITH}, 
numbers=left, 
numbersep=3pt, 
numberstyle=\tiny\color[rgb]{0.5,0.5,0.5}, 
rulecolor=\color{black}, 
showstringspaces=false, 
showtabs=false, 
stepnumber=0, 
tabsize=2, 
backgroundcolor=\color{white},
}
\lstdefinestyle{other}{ 
language=python,
basicstyle=\scriptsize\ttfamily, 
breakatwhitespace=false, 
breaklines=true, 
captionpos=b, 
commentstyle=\itshape\color[rgb]{0.4,0.4,0.4}, 
deletekeywords={VALUE, INPUT}, 
firstnumber=1, 
frame=none, 
frameround=tttt, 
keywordstyle=\color[RGB]{0, 110, 184},
morekeywords={}, 
numbers=left, 
numbersep=3pt, 
numberstyle=\tiny\color[rgb]{0.5,0.5,0.5}, 
rulecolor=\color{black}, 
showstringspaces=false, 
showtabs=false, 
stepnumber=0, 
tabsize=2, 
backgroundcolor=\color{white},
}
\lstdefinestyle{coq}{ 
language=python,
basicstyle=\scriptsize\ttfamily, 
breakatwhitespace=false, 
breaklines=true, 
captionpos=b, 
commentstyle=\itshape\color[rgb]{0.4,0.4,0.4}, 
deletekeywords={}, 
firstnumber=1, 
frame=none, 
frameround=tttt, 
keywordstyle=\color[RGB]{0, 110, 184},
morekeywords={Proof, Qed}, 
numbers=left, 
numbersep=3pt, 
numberstyle=\tiny\color[rgb]{0.5,0.5,0.5}, 
rulecolor=\color{black}, 
showstringspaces=false, 
showtabs=false, 
stepnumber=0, 
tabsize=2, 
backgroundcolor=\color{white},
}
\newcommand{\lbr}{\llbracket}
\newcommand{\rbr}{\rrbracket}

\newcommand{\teq}{\triangleq}

\newcommand{\set}[1]{\{#1\}}                    
\newcommand{\setof}[2]{\{{#1}\mid{#2}\}}        

\newcommand{\Integer}{\mathds{Z}}

\newcommand{\Bool}{\mathds{B}}
\newcommand{\true}{\texttt{true}}

\newcommand{\Query}{Query}
\newcommand{\Pred}{Predicate}
\newcommand{\Expr}{Expression}
\newcommand{\Proj}{Projection}

\newcommand{\lambdaFn}[2]{\lambda \; {#1} \;.\;{#2}}
\renewcommand{\mathtt}[1]{{\textrm {\tt {#1}}}}
\newcommand{\D}[1]{\lbr{#1}\rbr}

\newcommand{\N}{\mathds{N}}
\newcommand{\denote}[1]{\lbr #1 \rbr}
\newcommand{\sep}{~|~}
\newcommand{\Context}{\Gamma}
\newcommand{\context}{g}
\newcommand{\Schema}{\texttt{Schema}}
\newcommand{\schema}{\sigma}
\newcommand{\tuple}{t}
\newcommand{\query}{q}
\newcommand{\proj}{p}
\newcommand{\pred}{b}
\newcommand{\metapred}{\beta}
\newcommand{\expr}{e}

\newcommand{\zero}{\textbf{0}}
\newcommand{\one}{\textbf{1}}

\newcommand{\leaf}[1]{\texttt{leaf} \; #1}
\newcommand{\node}[2]{\texttt{node} \; #1 \; #2}
\newcommand{\emptySchema}{\texttt{empty}}

\newcommand{\Attr}{\texttt{Attribute}}
\newcommand{\agg}{\texttt{agg}}

\newcommand{\denoteQuery}[3]{\denote{#1 \vdash #2 : #3}}
\newcommand{\denoteExpr}[3]{\denote{#1 \vdash #2 : #3}}
\newcommand{\denoteProj}[3]{\denote{#1 : #2 \Rightarrow #3}}
\newcommand{\denotePred}[2]{\denote{#1 \vdash #2}}
\newcommand{\denoteTable}[1]{\denote{#1}}

\newcommand{\intro}{\lambda \; \context. \;}
\newcommand{\introt}{\lambda \; \tuple. \;}
\newcommand{\intros}{\lambda \; \context \; \tuple. \;}
\newcommand{\callQuery}[3]{\denoteQuery{#1}{#2}{#3} \; \context \; \tuple}
\newcommand{\callExpr}[3]{\denoteExpr{#1}{#2}{#3} \; \context}
\newcommand{\callPred}[2]{\denotePred{#1}{#2} \; \context}

\newcommand{\Tuple}{\texttt{Tuple}}
\newcommand{\Type}{\mathcal{U}}

\newcommand{\type}{\tau}

\newcommand{\mkPair}[2]{(#1,#2)}
\newcommand{\fst}[1]{{#1}.\texttt{1}}
\newcommand{\snd}[1]{{#1}.\texttt{2}}

\newcommand{\Unit}{\texttt{Unit}}

\newcommand{\merely}[1]{\left\lVert#1\right\rVert}

\newcommand{\unot}[1]{\texttt{not}(#1)}

\newcommand{\SELECT}[2]{\texttt{SELECT} \; #1 \; #2}
\newcommand{\WHERE}[2]{#1 \; \texttt{WHERE} \; #2}
\newcommand{\FROM}[1]{\texttt{FROM} \; #1}
\newcommand{\UNIONALL}[2]{#1 \; \texttt{UNION ALL} \; #2}
\newcommand{\EXCEPT}[2]{#1 \; \texttt{EXCEPT} \; #2}
\newcommand{\AND}[2]{#1 \; \texttt{AND} \; #2}
\newcommand{\OR}[2]{#1 \; \texttt{OR} \; #2}
\newcommand{\CastPred}[2]{\texttt{CASTPRED} \; #1 \; #2}
\newcommand{\CastExpr}[2]{\texttt{CASTEXPR} \; #1 \; #2}
\newcommand{\Compose}[2]{ #1 . \; #2}
\newcommand{\Duplicate}[2]{#1 , \; #2}
\newcommand{\DISTINCT}[1]{\texttt{DISTINCT} \; #1}
\newcommand{\NOT}[1]{\texttt{NOT} \; #1}
\newcommand{\EXISTS}[1]{\texttt{EXISTS} \; #1}

\newcommand{\Evaluate}[1]{\texttt{E2P}~#1}
\newcommand{\TRUE}{\texttt{TRUE}}
\newcommand{\FALSE}{\texttt{FALSE}}
\newcommand{\ProjL}{\texttt{Left}}
\newcommand{\ProjR}{\texttt{Right}}
\newcommand{\Empty}{\texttt{Empty}}
\newcommand{\Star}{\texttt{*}}
\newcommand{\Var}[1]{\texttt{P2E}~#1}

\newcommand{\SelectFrom}[2]{\SELECT{#1}{\FROM{#2}}}
\newcommand{\SelectFromWhere}[3]{\SelectFrom{#1}{\WHERE{#2}{#3}}}

\newtheorem{theorem}{Theorem}[section]

\newtheorem{definition}[theorem]{Definition}

\newtheorem{example}[theorem]{Example}
\newtheorem{lemma}[theorem]{Lemma}

\newcommand{\sqlString}{\texttt{string}}
\newcommand{\sqlBool}{\texttt{bool}}
\newcommand{\sqlInt}{\texttt{int}}
\newcommand{\attr}[2]{{#1}.\texttt{#2}}
\newcommand{\rel}[1]{\texttt{#1}}
\renewcommand{\paragraph}[1]{\vspace{2mm} \noindent \textbf{#1}}

\newcommand{\trans}[2]{\texttt{Tr}_{#1}(#2)}
\newcommand{\rsa}{\rightsquigarrow}
\newcommand{\hcons}[2]{#1 \; \bigoplus \; #2}
\newcommand{\ct}[2]{\texttt{Ct}_{#1}(#2)}
\newcommand{\lkup}[2]{\texttt{topath}(#1, #2)}
\newcommand{\sof}[1]{\texttt{table\_schema}(#1)}
\newcommand{\sofc}[2]{\texttt{alias\_schema}(#1, #2)}
\newcommand{\typeof}[2]{\texttt{typeof}(#1, #2)}
\newcommand{\rewrite}{\rightsquigarrow}

\let\oldequiv\equiv
\renewcommand{\equiv}{\textcolor{cyan}{\oldequiv}}
\newcommand{\eq}{=}
\newcommand{\red}[1]{\textcolor{red}{#1}}

\newcommand\sem{\textsc{SQL}\xspace}

\newcommand\outputLang{\textsc{U-expression}\xspace}
\newcommand\hottir{\textsl{SQL}\textsc{IR}\xspace}
\newcommand\newexp{U-expression\xspace}
\newcommand\newexps{U-expressions\xspace}
\newcommand\firstnf{\textsc{SPNF}\xspace}

\newcommand\algname{\texttt{UDP}\xspace}
\newcommand\setalgname{\texttt{SDP}\xspace}
\newcommand\termalgname{\texttt{TDP}\xspace}

\newcommand{\defeq}{\stackrel{\text{def}}{=}}

\newcommand{\sys}{Cosette}

\newcommand{\reviewermeta}[1]{\textcolor{black}{#1}}
\newcommand{\reviewerone}[1]{\textcolor{black}{#1}}
\newcommand{\reviewertwo}[1]{\textcolor{black}{#1}}
\newcommand{\reviewerthree}[1]{\textcolor{black}{#1}}

\newcommand{\ourring}{U-semiring\xspace}
\newcommand{\ourrings}{U-semirings\xspace}
\newcommand{\oldsys}{\textsc{Cosette}\xspace}

\newcommand{\figlabel}[1]{\label{fig:#1}}
\newcommand{\seclabel}[1]{\label{sec:#1}}

\newcommand{\rlabel}[1]{\label{rule:#1}}

\newcommand{\secref}[1]{Sec.~\ref{sec:#1}}  
\newcommand{\figref}[1]{Fig.~\ref{fig:#1}}     

\newcommand{\rref}[1]{Rule~(\ref{rule:#1})}

\setlist[itemize]{itemsep=0mm, leftmargin=*}
\setlist[enumerate]{itemsep=1mm}
\def\sharedaffiliation{%
\end{tabular}
\begin{tabular}{c}}

\vldbTitle{}
\vldbAuthors{}
\vldbDOI{}

\begin{document}
\title{Axiomatic Foundations and Algorithms for Deciding Semantic Equivalences of SQL Queries}

\numberofauthors{1} 

\author{
  \alignauthor Shumo Chu, Brendan Murphy, Jared Roesch, Alvin Cheung, Dan Suciu\\
  \sharedaffiliation
  \affaddr{Paul G. Allen School of Computer Science and Engineering} \\
  \affaddr{University of Washington} \\
  \affaddr{{\tt \{chushumo, jroesch, akcheung, suciu\}@cs.washington.edu}, \tt bsmurphy@uw.edu}
}

%
%


\maketitle

\begin{abstract}

Deciding the equivalence of SQL queries is a fundamental problem in data management. As prior work has mainly focused on studying the theoretical limitations of the problem, very few implementations for checking such equivalences exist. 
In this paper, we present a new formalism and implementation for reasoning about the equivalences of SQL queries. Our formalism, \ourring, extends SQL's semiring semantics with unbounded summation and duplicate elimination. \ourring is defined using only very few axioms and can thus be easily implemented using proof assistants such as Coq for automated query reasoning. Yet, they are sufficient enough to enable us reason about sophisticated SQL queries that are evaluated over bags and sets, along with various integrity constraints.
To evaluate the effectiveness of \ourring, we have used it to formally verify \reviewermeta{68 equivalent queries and} rewrite rules from both classical data management research papers and real-world SQL engines, where many of them have never been proven correct before.


\end{abstract}


\section{Introduction}
\label{sec:intro}


All modern relational database management systems (DBMSs) contain a
query optimizer that chooses the best means to execute an input
query. At the heart of an optimizer is a rule-based rewrite system
that uses rewrite rules to transform the input SQL
query into another (hopefully more efficient) query to 
execute. A key challenge in query optimization is how to
ensure that the rewritten query is indeed semantically equivalent to
the input, i.e., that the original and rewritten queries return the
same results when executed on all possible input database instances.
History suggests that the lack of tools to establish 
such equivalences has caused longstanding, latent bugs in major database
systems~\cite{countBug}, and new bugs continue to arise
\cite{pgbug, oraclebug}. All such errors lead to incorrect results 
returned that can cause dire consequences.

In the past, query optimizers were
developed by only small number of commercial database vendors with dedicated
teams to handle customer reported bugs. Such teams effectively acted
as ``manual solvers'' for query equivalences. The explosive
growth of new data analytic systems in recent decades (e.g., Spark SQL \cite{sparksql}, Hive~\cite{hive}, Dremel~\cite{dremel}, Myria~\cite{myria1, myria2}, among many others) has unfortunately exacerbated this problem significantly, as many such new systems are developed by teams that simply lack the dedicated resources to check every single rewrite implemented in their system. 
For example, the Calcite open source query processing engine~\cite{calcite} is mainly developed by open-source contributors. While it contains more than $232$ rewrite rules in its optimization engine, none of which has been formally validated. The same holds true for similar engines.



On the other hand, 
deciding whether two arbitrary SQL queries are semantically equivalent
is a well-studied problem in the data management
research community that has shown to be undecidable in
general~\cite{AbiteboulHV95}. Most subsequent research has been
directed at identifying fragments of SQL where equivalence is
decidable, under set semantics~\cite{cqDecidable, SagivY80JACM} or
bag semantics~\cite{CohenNS99}. This line of work, focused on
theoretical aspects of the problem \cite{Trakhtenbrot50DANUSSR,
 SagivY80JACM, JayramKV06PODS}, led to very few implementations,
most of which were restricted to applying the chase procedure to conjunctive
queries~\cite{benediktSurvey}.

An alternative approach to query equivalence was recently proposed by the
{\oldsys} system~\cite{cosette_pldi} based on a new SQL
semantics. It interprets SQL relations as
$K$-relations~\cite{GreenKT07PODS}. 
A $K$-relation is semiring that maps each tuple $t$ to a value $R(t)$
that denotes the multiplicity of the tuple in the relation.  Normally, the
multiplicity is an integer, thus the semiring $K$ is the semiring of
natural numbers $\N$, but in order to support some of the SQL
constructs (such as projections, with or without \texttt{DISTINCT}), {\oldsys} resorts to using a much more complex semiring, namely that of all univalent types in Homotopy Type Theory (HoTT)~\cite{hottbook}.  Equivalence of
two SQL queries is then proven in {\oldsys} using the Coq proof assistant
extended with the HoTT library~\cite{HoTTCoq}.  While the system can
handle many features of SQL, its reliance on HoTT makes it difficult
to extend as HoTT is both a Coq library under development and 
an active research area itself. 
As a result, {\oldsys} has a number of shortcomings. 
For example, {\oldsys} does not support foreign
keys as they are difficult to model using the HoTT library in Coq. 
In this paper, we propose a new algebraic structure, called 
the \emph{unbounded semiring}, or \emph{{\ourring}}. 
We define the \ourring by extending the standard commutative 
semiring with a few simple constructs and axioms. 
This new algebraic structure 
serves as a foundation for our new formalism that models the semantics of SQL. To prove the semantic equivalence of two SQL queries, we first convert them into {\ourring} expressions, i.e., \emph{{\newexps}}.
Deciding the equivalence of queries then becomes determining the equivalence of two {\newexps} which, as we will show, is much easier compared to classical approaches. 

Our core contribution is identifying the minimal set of axioms for \ourrings that are sufficient to prove sophisticated SQL query equivalences. This is important as the number of axioms determines the size of the trusted code base in any proof system.
As we will see, the few axioms we designed for \ourrings are surprisingly simple as they are just identities between two \newexps. Yet, they are sufficient to prove various advanced query optimization rules that arise in real-world optimizers.
Furthermore, we show how integrity constraints (ICs) such as keys and foreign keys can be expressed as \newexps identities as well, and this leads to a single framework that can model many different features of SQL and the relational data model.
This allows us to devise different algorithms for deciding the equivalences of various types of SQL queries including those that involve views, or leverage ICs to rewrite the input SQL query as part of the proof.

To that end, we have developed a new algorithm, \algname, to automatically decide 
the equivalence of two arbitrary SQL queries that are 
evaluated under mixed set/bag semantics,\footnote{mixed set/bag semantics is the bag semantics that allows explicit \texttt{DISTINCT} on arbitrary subqueries, bag semantics and set semantics are special cases of mixed set/bag semantics.} and also in the presence of indexes, views, and other ICs. 
At a high level, our algorithm performs rewrites reminiscent to the chase/back-chase 
procedure~\cite{PopaDST00} but uses \newexps 
rather than first order logic sentences. It performs
a number of tests for isomorphisms or homomorphisms to capture the
mixed set/bag semantics of SQL. 
Our algorithm is sound in general and is complete in two restricted cases: 
when the two queries are Unions of Conjunctive Queries (UCQ) under set semantics, 
or UCQ under bag semantics. 
We implement \algname on top of \reviewermeta{the Lean proof assistant} \cite{lean}. 
The implementation includes the modeling of relations and queries as \newexps, 
axiomatic representations of ICs as simple \newexp identities, and 
the algorithm for checking the equivalence of \newexps. 

We evaluate \algname using various optimization rules 
from classical data management research literature and 
as implemented in the Apache Calcite framework~\cite{calcite}. 
These rules consist of sophisticated SQL queries with 
a wide range of features such as subqueries, 
grouping and aggregate, \texttt{DISTINCT}, and integrity constraints.In fact, only 1 of them have been proven before. 
\algname can formally and automatically prove most of them (39 of 45). The running time of \algname on each of these 39 rules is within 30 seconds.
\reviewerone{Our system only proves {\em equivalence}. In
  prior work~\cite{cosette_cidr} we described the complementary
  task that uses a model checker to find counterexamples to identify buggy rewrites~\cite{countBug,pgbug,oraclebug}; of course, the model checker
  cannot prove equivalence, which is our current focus.}

In summary, our paper makes the following contributions:
\begin{itemize}
\item We describe a new algebraic structure, the \ourring, which extends the standard semiring with the necessary operators to model the semantics of a wide range of SQL queries (\secref{axioms}). 

\item We propose a new formalism for expressing different kinds of
  integrity constraints over a \ourring. We implement such constraints
  using a number of axioms (in the form of simple identities) such that they can be easily implemented using a proof assistant. 
  Doing so allows us to easily utilize them in equivalence proofs (Sec.~\ref{sec:constraints}).

\item We describe a new algorithm for deciding the equivalences of
  different types of SQL queries. The algorithm operates entirely on
  our representation of SQL queries as \newexps. We show that this algorithm is
  sound for general SQL queries and is complete for Unions of Conjunctive Queries under set and bag semantics (\secref{alg}).


\item We have implemented \algname using the Lean proof assistant, and have evaluated \algname by collecting \reviewerthree{69} real-world rewrite rules as benchmarks, 
both from prior research work done in the database research community, and from Apache Calcite~\cite{calcite}, 
an open-source relational query optimizer.
To our knowledge, the majority of these rules have been never proven correct before, while  
\algname can automatically prove \reviewerthree{62} of \reviewerthree{68} correct ones 
and fail as intended on the 1 buggy one (\secref{evaluation}). 


\end{itemize}



\section{Overview}

\label{sec:overview}

\begin{figure}
\begin{small}
  \begin{center}
\textbf{Equivalent SQL Queries}
\begin{lstlisting}[style=sql]
SELECT * FROM R t WHERE t.a >= 12  -- Q1

SELECT t2.* FROM I t1, R t2        -- Q2
WHERE t1.k = t2.k AND t1.a >= 12
\end{lstlisting}
where \texttt{k} is a key of \texttt{R}, and {\tt I} is an index on {\tt R} defined as:\\
\begin{lstlisting}[style=sql]
I := SELECT t3.k AS k, t3.a AS a FROM R t3
\end{lstlisting}
\end{center}
\vspace{-0.3in}
\begin{center}
  \textbf{Corresponding Equivalence in Semirings}
  \begin{flalign*}
    &Q_1(t) \eq \lambda\;t.\; \denote{R}(t) \times [t.a \geq 12]  \\
    & Q_2(t) \eq \lambda\;t.\; \sum_{t_1, t_2, t_3} [t_2 = t] \times [t_1.k =
    t_2.k] \times [t_1.a \geq 12] \times
    \\
    & \qquad [t_3.k = t_1.k] \times[t_3.a = t_1.a] \times
    \denote{R}(t_3) \times \denote{R}(t_2)
  \end{flalign*}
\end{center}
\end{small}
\vspace{-0.2in}
\caption{Proving that a query is equivalent to a rewrite using an index
  $I$ requires proving a subtle identity in a semiring.
}
\figlabel{index-rewrite}
\vspace{-3mm}
\end{figure}

We motivate our new semantics using an example query rewrite. As shown
in \figref{index-rewrite}, the rewrite changes the original query {\tt
  Q1} by using an index $I$ for look up.  We follow the GMAP
framework~\cite{TsatalosSI96}, where an index is considered as a view
definition, and a query plan that uses the index $I$ to access the
relation $R$ is represented logically by a query that selects from $I$
then joins on the key of $R$.

Our goal is to devise a semantics for SQL along with various integrity constraints that can be easily implemented as a tool for checking query equivalences.
Unfortunately, the SQL standard~\cite{Date89AW} is expressed in English and is difficult to implement programmatically. One recent attempt is Q*cert~\cite{AuerbachHMSS17SIGMOD}, which models SQL using
NRA (Nested Relational Algebra). NRA is implemented using the Coq proof assistant, with relations modeled using lists.
To prove that $Q_1$ and $Q_2$ are equivalent in Q*cert, 
users need to prove that the output from the two are equal up to
element reordering and duplicate elimination (for set semantics).
As Q*cert models relations as lists, this amounts to writing an inductive proof on the size of $R$, i.e., if $R$ is empty, then $Q_1$ outputs the same relation as $Q_2$; if $R$ is of size $n$ and the two outputs are equivalent, then the two outputs are also equivalent if $R$ is of size $n+1$. Writing such proofs can be tedious.
As an illustration, Q*cert requires $45$ 
lines of Coq to prove that selection can be distributed over union \cite{qcert_proof} while, as we will see, using \ourring only requires $1$ line of Coq: distribute multiplication over addition, which is one of the semiring axioms.

Another formalism proposed by Guagliardo et al.~\cite{GuagliardoL17} models relations as bags. While it models {\tt NULL} semantics, it (like Q*cert) does not model integrity constraints, which are widely used in almost all database systems and are involved in the rewrites in many real world optimizers. There is no known algorithm based on their formalism to automatically check the equivalence of SQL queries.

We instead base our semantics on K-relations, which was first proposed by Green et
al~\cite{GreenKT07PODS}. Under this semantics, a relation is modeled
as a function that maps tuples to a commutative semiring,
${\bf K} = (K, 0, 1, +, \times)$. In other words, a K-relation, $R$,
is a function:
%
%
\begin{align*}
  \denote{R}: & \Tuple(\sigma)  \rightarrow K
\end{align*}
with finite support.  Here, $\Tuple(\sigma)$ denotes the (possibly
infinite) set of tuples of type $\sigma$, and $\denote{R}(t)$
represents the multiplicity of $t$ in relation $R$.  For example, a
relation under SQL's standard bag semantics is an $\N$-relation
(where $\N$ is the semiring of natural numbers), and a relation under set
semantics is a $\Bool$-relation (where $\Bool$ is the semiring of Booleans).
All relational operators and SQL queries can be expressed in terms of
semiring operations; for example:
%
\begin{align} 
  & \denote{\SelectFrom{\Star}{\texttt{$R$ $x$, $S$ $y$}}} =  \lambdaFn{(t_1,t_2)}{\denote{R}(t_1) \times \denote{S}(t_2)} \nonumber \\
  & \denote{\SelectFromWhere{\Star}{\texttt{$R$}}{a>10}} =  \lambdaFn{t}{[t.a>10] \times \denote{R}(t)}\nonumber \\
  & \denote{\texttt{SELECT $a$ FROM $R$}} = \introt \sum_{\tuple': \texttt{Tuple}(\sigma)}  [t'.a = t] \times \denote{R}(t') \nonumber 
\end{align}
%
\noindent For any predicate $\pred$, 
we denote $[\pred]$ the element of the semiring defined by $[\pred] = 1$ if the
predicate holds, and $[\pred] = 0$ otherwise. 

$K$-relations can be used to prove many simple query rewrites by reducing 
query equivalences to semiring equivalences, which are much easier since 
one can use algebraic reasoning rather than proofs by induction.  
However, for sophisticated rewrites like the one shown in \figref{index-rewrite}, 
$K$-relations are not sufficient to prove query equivalences, 
let alone automate the proof search. 
One major problem is that $K$-relation didn't model integrity constraints 
(keys, foreign keys, etc.), which are usually expressed as generalized
dependencies~\cite{AbiteboulHV95} (logical formulas of the
form $\forall {\mathbf{x}} (\varphi \Rightarrow \exists {\mathbf{y}}
\psi)$).
For example, consider the two queries in
Fig.~\ref{fig:index-rewrite}: $Q_1$ is a selection on
$R.a \geq 12$, while $Q_2$ rewrites the given query using an
index $I$.  These two queries are indeed
semantically equivalent in that scanning a 
table using a given attribute ($a$ in the example) is the same as
performing an index scan on the same attribute.
However, consider their corresponding expressions over $K$-relations,
shown at the bottom of Fig.~\ref{fig:index-rewrite}; 
it is unclear how to express formally the fact that $R.k$ is a key, yet alone prove automatically that these two expressions are equal.

To extend $K$-relations for \emph{automatically} proving SQL equivalences 
under database integrity constraints, we develop a novel algebraic structure,
\ourring, as the semiring in $K$-relations.
A \ourring extends a semiring with 
three new operators, $\sum$, $\merely{\cdot}$, $\unot{\cdot}$, and a minimal
set of axioms, each of which is a simple identity\footnote{\small An {\em axiom} is
a logical sentence, such as $\forall x R(x) \Rightarrow S(x)$.  An
  {\em identity}, or an {\em equational law}, is an equality, such as
  $x+y=y+x$.  The implication problem for identities is much simpler
  than for arbitrary axioms.  Traditional generalized
  dependencies~\cite{AbiteboulHV95} are expressed as axioms,
  $\forall {\mathbf{x}} (\varphi \Rightarrow \exists {\mathbf{y}}
  \psi)$,
  and only apply to queries under set semantics.} 
that is easy to implement using a proof assistant.
We also discover a set of \ourring identities that models 
database integrity 
constraints such as keys and foreign keys.
With these additions, SQL equivalences can be reasoned by checking the equivalences of their
corresponding {\newexp}s using only these axioms 
expressed in \ourring identities. 

Using our semantics, the rewrite shown in \figref{index-rewrite} can be
proved by rewriting the $Q_2$ into the $Q_1$ using \ourring axioms in three steps. 
First, the sum over $t_1$ can be removed by applying the axiom of the interpretation of equality over summation (Eq.~\eqref{eq:sum-subst-eq}):
\begin{flalign*}
    \lambda \; t. \; \sum_{t_2,t_3} [t_2 = t] \times [t_3.k = t_2.k] \times [t_3.a \geq 12] \times R(t_3) \times R(t_2)
  \end{flalign*}
Since $k$ is a key of $R$, applying the \ourring definition of the key constraint (Def.~\ref{def:key}) we get:
\[ [t_3.k = t_2.k] \times  \denote{R}(t_3) \times \denote{R}(t_2) \eq  
[t_3 = t_2] \times \denote{R}(t_3) \]
Thus, the $Q_2$ can be rewritten to:
\[ \lambda \; t. \; \sum_{t_2,t_3} [t_2 = t] \times [t_3.a \geq 12] \times \denote{R}(t_2) \times [t_2 = t_3] \]
Applying Eq.~\eqref{eq:sum-subst-eq} again to the above makes $Q_2$ is equivalent to $Q_1$. We present a detailed proof in Ex~\ref{ex-index}.

We next explain these axioms in detail and show how we use them to develop an algorithm to 
formally and automatically check the equivalences of general SQL queries.

\section{Axiomatic Foundations}
\seclabel{axioms}
\begin{figure}[t]
\centering
\begin{small}
\[
\begin{array}{llll}
  h \in \texttt{Program} & ::= & s_1; \ldots; s_n; \\
  s \in \texttt{Statement} & ::= & \texttt{verify} \; q_1 \oldequiv q_2 \\
    & \; \mid & \texttt{schema} \; \schema(a_1:\type_1, \ldots, a_n:\type_n)\\
    & \; \mid & \texttt{table} \; r(\schema) \\
    & \; \mid & \texttt{key} \; r(a_1, \ldots, a_n); \\
    & \; \mid & \texttt{foreign key} \; r_1(a_1', \ldots, a_n') \; \\
    & & \ \ \ \ \ \texttt{references} \; r_2(a_1, \ldots, a_n); \\
    & \; \mid & \texttt{view} \; v \;q; \\
    & \; \mid & \texttt{index} \; i \; \texttt{on} \; \schema(a_1, \ldots, a_n); \\
  a \in \texttt{Attribute} & ::= & \text{string} \\
  q \in \texttt{Query} & ::=  &  r \\
        & \; \mid & \texttt{SELECT} \; p \; q      \\
        & \; \mid & \texttt{FROM} \; q_1 \; x_1, \ldots, q_n \; x_n     \\
        & \; \mid & q \; \texttt{WHERE} \; p        \\
        & \; \mid & q_1 \; \texttt{UNION ALL} \; q_2   \\
        & \; \mid & \texttt{DISTINCT} \; q \\
 x \in \texttt{TableAlias} & ::= & \text{string}  \\
 b \in \texttt{Predicate} & ::= & e_1 \; \texttt{=} \; e_2 \\
       & \; \mid &  \texttt{NOT} \; b
            \mid b_1 \; \texttt{AND} \; b_2
            \mid b_1 \; \texttt{OR} \; b_2 \\
        & \; \mid & \texttt{TRUE} \mid \texttt{FALSE} \\
       & \; \mid & \texttt{EXISTS} \; q \\
 e \in \texttt{Expression} & ::=  & x.a \mid f(e_1, \ldots, e_n) \mid \agg(q)   \\
 p \in \texttt{Projection} & ::= & \Star \mid x.\Star \mid e \; \texttt{AS} \; a \mid \Duplicate{p_1}{p_2} \\
 f \in \texttt{UDF}, \agg \in \texttt{UDA} & ::= & \text{string}
\end{array}
\]
\end{small}
\vspace{-0.2in}
\caption{SQL fragment supported by our semantics}
\figlabel{sql-syntax}
\vspace{-0.1in}
\end{figure}

In this section we introduce a new algebraic structure, \ourring, and
show that it can be used to give semantics to SQL queries.




\subsection{{\ourring}s}
\seclabel{ring}

Under standard bag semantics, a SQL query and its input relations can
be modeled as $\N$-relations~\cite{GreenKT07PODS}, i.e., relations
over the semiring $(\N, 0, 1, +, \times)$.  However, as we saw SQL
queries include constructs that are not directly expressible in a
semiring, such as projection (requiring an unbounded summation), {\tt
DISTINCT}, and non-monotone operators (e.g., {\tt NOT EXISTS}). We now define a new algebraic structure that can be used for such purposes.


\begin{definition}
  \label{ourring-def} An {\em unbounded semiring}, or {\ourring}, is
  $(\Type, \zero, \one, +, \times, \merely{\cdot}, \unot{\cdot}, (\sum_D)_{D \in \mathcal{D}})$, where:
\begin{itemize}
\item $(\Type, \zero, \one, +, \times)$ forms a commutative
  semiring.\footnote{\small Recall that the semiring axioms are:
    associativity and commutativity of $+$ and $\times$, identity of
    $\zero$ for $+$, identity of $\one$ for $\times$, distributivity
    of $\times$ over $+$, and $\zero \times x = \zero$;
    see~\cite{GreenKT07PODS} for details.}
\item $\merely{\cdot}$ is a unary operation called {\em squash}
  that satisfies:
\begin{eqnarray}
  \merely{\zero}=\zero &,& \reviewermeta{\merely{\one + x} = \one} \label{eq:merely-zero-one} \\
  \merely{\merely{x}+y} &=& \merely{x+y} \label{eq:merely-plus} \\
  \merely{x} \times \merely{y} &=& \merely{x \times y} \label{eq:merely-times} \\
  \merely{x} \times \merely{x} &=& \merely{x} \label{eq:merely-square} \\
 \reviewermeta{x \times \merely{x}} &\reviewermeta{=}& \reviewermeta{x} \label{eq:time-merely-keeps} \\
  x^2 = x &\Rightarrow& \merely{x} = x \label{eq:square-merely}
\end{eqnarray}
%
\item $\unot{\cdot}$ is a unary operation that satisfies the following:
  \begin{eqnarray*}
    \unot{\zero} &=& \one \\
    \unot{x \times y} &=&  \merely{\unot{x} + \unot{y}} \\
    \unot{x + y} &=& \unot{x}  \times \unot{y} \\
    \unot{\merely{x}} &=& \merely{\unot{x}} = \unot{x}
  \end{eqnarray*}
\item $\mathcal{D}$ is a set of sets; each $D \in \mathcal{D}$ is
  called a {\em summation domain}. For each $D$, the operation
  $\sum_D : (D \rightarrow \Type) \rightarrow \Type$ is called an {\em
    unbounded summation}: its input is a function
  $f: D \rightarrow \Type$, and its output is a value in $\Type$.
  We will write the summation as $\sum_{t \in D} f(t)$, or just
  $\sum_t f(t)$, where $f$ is an expression and $t$ is a free
  variable.
\item Unbounded summation further satisfies the following axioms:
  \begin{eqnarray}
    \sum_{t}\big(f_1(t) + f_2(t)\big) &=& \sum_{t}f_1(t) + \sum_{t}f_2(t) \label{eq:sum-distr} \\
    \sum_{t_1} \sum_{t_2} f(t_1, t_2) &=& \sum_{t_2}\sum_{t_1} f(t_1, t_2) \label{eq:sum-symm} \\
 x \times \sum_{t} f_2(t) &=& \sum_{t} x \times f_2(t) \label{eq:prod-sum-assoc} \\
 \reviewermeta{\merely{\sum_{t} f(t)}} & \reviewermeta{=} & \reviewermeta{\merely{\sum_{t} \merely{f(t)}} \label{eq:squash-sum-squash}}  
  \end{eqnarray}
%
\end{itemize}
\end{definition}

Thus, a {\ourring} extends the semiring with unbounded summation, the
squash operator (to model the SQL {\tt DISTINCT} operator), and
negation (to model {\tt NOT EXISTS}).
We chose a set of axioms that captures the semantics of SQL queries.
For example, Eq.\eqref{eq:merely-plus} implies $\merely{\merely{x}}=\merely{x}$,
which, as we will see, helps us prove that \texttt{DISTINCT} of \texttt{DISTINCT}
equals \texttt{DISTINCT}.  Eq.\eqref{eq:merely-square} captures
equality of queries under set semantics, such as
\texttt{SELECT DISTINCT R.a FROM R}
and \texttt{SELECT DISTINCT x.a FROM R x, R y WHERE x.a = y.a}.
Eq.\eqref{eq:square-merely}
captures even more subtle interactions between subqueries with and
without \texttt{DISTINCT} as described in~\cite{PiraheshHH92}. We will
illustrate this in~\secref{alg-example}.

If a summation domain $D$ is finite, say $D = \set{a_1, \ldots, a_n}$,
we could define $\Sigma_D f$ directly as
$f(a_1) + \dots + f(a_n)$.  However, as we saw, the meaning of a
projection requires us to sum over an unspecified set, namely, all
tuples of a fixed schema. Even though all SQL summation domains
are finite, {\em proving} this automatically is difficult and adds
considerable complexity even for the simplest query equivalences (for instance, we need to prove that all operators preserve domain finiteness). Instead,
we retain unbounded summation as a primitive in a {\ourring}.

We now illustrate four simple examples of \ourrings.  (1) If all summation
domains are finite, then the set of natural numbers, $\N$, forms a
\ourring, where the unbounded summation is the standard sum, the
squash and negation operators are $\merely{0}=\unot{x}=0$, and
$\merely{x}=\unot{0}=1$ for all $x\neq 0$.  (2) Its {\em closure},
$\bar \N \defeq \N \cup \set{\infty}$, forms a \ourring\ over arbitrary
summation domains.\footnote{\small $+$ and $\times$ are extended by
  $x + \infty = \infty$, $0 \times \infty = 0$, and
  $x \times \infty = \infty$ for $x\neq 0$.  Unbounded summation is
  defined as $\sum_D f \defeq f(a_1)+\cdots+f(a_n)$ when the support
  of $f$ is finite,
  $\texttt{supp}(f)\defeq \setof{x}{f(x) \neq 0} =
  \set{a_1,\ldots,a_n}$,
  and $\sum_D f = \infty$ otherwise.}  (3) The univalent
types~\cite{hottbook} form an \ourring; in our prior system,
\oldsys~\cite{cosette_pldi}, we used univalent types to prove SQL
query equivalence.  (4) Finally, the cardinal numbers (which form a
subset of univalent types) also form a {\ourring}.

To appreciate the design of \ourrings, it may be helpful to see what we
\emph{excluded}. Recall that fewer axioms translate into a
simpler proof system; hence, the need for frugality.
A {\em complete-semiring}~\cite{complete-semirings} also extends a
semiring with unbounded summation. However, it requires a stronger set
of axioms: summation must be defined over {\em all} subsets of some index
set, and includes additional axioms, such as
$\sum_{\set{a,b}} f = f(a)+f(b)$, among others. In a {\ourring},
summation is defined on only a small and fixed set of summation
domains that do not include $\set{a,b}$. This removes the need for
axioms involving complex index sets. Similarly, the axioms for
$\merely{\cdot}$ and $\unot{\cdot}$ are also kept to a minimum; for
example, we omitted unnecessary conditional identities like {\em if
  $x \neq \zero$ then $\merely{x} = \one$}.  One example of a
\ourring\ where this conditional axiom fails is the set of diagonal
$2 \times 2$ matrices with elements in $\bar \N$, where
$\zero \defeq \text{diag}(0,0)$, $\one \defeq \text{diag}(1,1)$, and
all operations are performed on the diagonal using their meaning in
$\bar \N$. In this semiring,
$\merely{x} \in \set{\text{diag}(0,0), \text{diag}(0,1),
 \text{diag}(1,0), \text{diag}(1,1)}$.

Another important decision was to exclude order relations, $x \leq y$,
which we could have used to define key constraints (by stating that
every key value occurs $\leq \one$ times).  An {\em ordered semiring}
(see also a {\em dioid}~\cite{semiring_book}) is a semiring equipped
with an order relation $\leq$ such that $\zero \leq a$ for all $a$,
and $x \leq y$ implies $x+z \leq y+z$. We did not require a \ourring\
to be ordered, instead we define key constraints using only the
existing axioms (Sec.~\ref{sec:constraints}).

With all these simplifications, one wonders if it is possible to prove
\emph{any} non-trivial SQL equivalences, let alone those in the presence of
complex integrity constraints. We answer this in the affirmative,
as we shall next explain.


\subsection{{\ourring} SQL Semantics}
\seclabel{denotation}

Every SQL query is translated into a \newexp, which denotes the
K-relation of the query's answer; to check the equivalences of two SQL
queries, the system first translates them to \newexps, then checks
their equivalence using the \algname algorithm described in
\secref{alg}.  In this section, we describe the translation from SQL
queries to \newexps.



\figref{sql-syntax} shows the SQL fragment currently supported by our
implementation.  We require the explicit declaration of table schemas,
keys, foreign keys, views, and indexes, and support a rich fragment of
SQL that includes subqueries and {\tt DISTINCT}.  We also support {\tt
  GROUP BY} by de-sugaring them into subqueries as follows:
\begin{lstlisting}[style=sql,basicstyle=\small\ttfamily]
SELECT x.k as k, agg(x.a) as a1 FROM R x
GROUP BY x.k
\end{lstlisting}
\vspace{-5mm}
\[\text{is rewritten to} \Downarrow \]
\vspace{-6mm}
\begin{lstlisting}[style=sql,basicstyle=\small\ttfamily, escapeinside={@}{@}]
SELECT @\textcolor{red}{y.k}@ as k,
       agg(SELECT x.a as a
           FROM R x WHERE x.k=y.k) as a1
FROM R y
\end{lstlisting}
\noindent where $R$ can be any SQL expression.
Our implementation of \algname currently supports aggregates by treating them
as uninterpreted functions.

Each SQL query $Q$ is translated into a {\em \newexp} as follows.  We
denote a \newexps\ by a capital letter $E$, denote a SQL expression by
a lower case letter $e$ ($\texttt{Expression}$ in \figref{sql-syntax},
e.g.,  $t.price/100$), and denote a SQL predicate by $b$
($\texttt{Predicate}$ in \figref{sql-syntax}, e.g.
$t.price/100 > s.discount$).


\begin{itemize}
\item For each schema $\sigma$ in the program, there is a summation
  domain $\Tuple(\sigma)$, which includes all tuples with schema $\sigma$.
\item For each relation name $R$, there is a predefined function
  $\denote{R}: \Tuple(\sigma) \rightarrow \Type$. Intuitively,
  $\denote{R}(\tuple)$ returns the multiplicity of $\tuple \in R$, or
  $\zero$ if $\tuple \not\in R$.  If the context is clear, we will drop
  $\denote{\cdot}$ and simply write $R(\tuple)$.  Unlike prior work~\cite{GreenKT07PODS}, we do not require the support of $R$ to
  be finite, as that would make it difficult to axiomatize as discussed.
\item For each SQL predicate expression $b$, there is a \newexp\
  $[b] \in \Type$ that satisfies:
  \begin{align}
    [b] = & \merely{[b]} \label{eq:pred-merely}
  \end{align}
  Under the standard interpretation of SQL, $[b]$ is either $\zero$ (false) or $\one$ (true),
  but formally we
  only require Eq.\eqref{eq:pred-merely}.  The equality predicate has
  a special interpretation in \ourring and is required to satisfy the
  following axioms, called {\em excluded middle} (Eq.~\eqref{eq:lem}), {\em substitution of
    equals by equals} (Eq.~\eqref{eq:subst-eq}), and {\em uniqueness of equality} (Eq.~\eqref{eq:sum-eq}):
  \begin{align}
    [e_1 = e_2] + [e_1 \neq e_2] = & \one \label{eq:lem}\\
    f(e_1) \times [e_1 = e_2]  = & f(e_2) \times [e_1 = e_2]\label{eq:subst-eq}\\
    \sum_t [t=e] = & \one \label{eq:sum-eq}
  \end{align}
  These axioms are sufficient to capture the semantics of equality ($=$). For example, we
  can prove:\footnote{\small Using (\ref{eq:subst-eq}), \ref{eq:prod-sum-assoc},
    and (\ref{eq:sum-eq}) we can show that:
    $\sum_t [t=e] \times f(t) = \sum_t [t=e] \times f(e) = f(e) \times
    \sum_t [t=e] = f(e) \times \one = f(e)$.}
 \begin{equation}
    \sum_t [t=e] \times f(t) = f(e) \label{eq:sum-subst-eq}
 \end{equation}
\item For each uninterpreted aggregate operator $\agg$ and \newexp\
  $E : \Tuple(\sigma) \rightarrow \Type$, $e ::= \agg(E)$ is a valid
  value expression.
It can be used in a predicate, e.g., $[\agg(E)=t.a]$.
\end{itemize}



Every SQL query $q$ is translated, inductively, to a \newexp\ denoted
$\denote{q}$.  For example, a table $R$ is translated into its
predefined semantics $\denote{R}$; a {\tt SELECT-FROM-WHERE} is
translated by generalizing $K$-relation SQL semantics (shown in
Sec.~\ref{sec:overview}); predicates are also translated inductively:
$\texttt{NOT}, \texttt{AND}, \texttt{OR}$ become
$\unot{\cdot}, \times, +$, while
$\denote{\texttt{EXISTS } q} = \merely{\denote{q}}$ and
$\denote{\texttt{NOT EXISTS } q} = \unot{\denote{q}}$.
\reviewerone{Notice that the language includes nested queries both in the
  {\tt FROM} and the {\tt WHERE} clauses; their translation is based
  on standard unnesting.} We omit the
details and state only the main property:
\reviewerthree{
\begin{definition}
  Every SQL expression $q$ in \figref{sql-syntax} is translated
  into a \newexp\, $\denote{q} : \Tuple(\sigma) \rightarrow \Type$.
  The translation is defined inductively on the structure of $q$;
  see the full paper for details~\cite{DBLP:journals/corr/abs-1802-02229}.
\end{definition}
}
We write $q$ instead of $\denote{q}$ when context permits.
Strictly speaking $q$ is a function,
$q = \lambda \tuple. E$, but we use the more friendly notation
$q(\tuple) = E(\tuple)$, where $E$ is an expression with a free
variable $\tuple$.


\subsection{Sum-Product Normal Form}
\seclabel{fnf}

\begin{figure}
\noindent \textbf{SQL Query $q$}
\begin{lstlisting}[style=sql,basicstyle=\footnotesize\ttfamily]
SELECT t2.*
FROM (SELECT t1.k as k, t1.a as a
      FROM R t3) t1, R t2
WHERE t1.k = t2.k AND t1.a $\geq$ 12
\end{lstlisting}
\noindent \textbf{Its corresponding  {\newexp} in {\firstnf}}
\begin{small}
\begin{flalign*}
\denote{q}(t) & =  \sum_{t_1,t_2} [t_2 = t] \times (\sum_{t_3} [t_3.k = t_1.k] \times [t_3.a = t_1.a] \times & \\
 & \qquad \; R(t_3)) \times R(t_2) \times [t_1.k = t_2.k] \times [t_1.a \geq 12] & \\
 \quad & \eq  \sum_{t_1,t_2} [t_2 = t] \times \textcolor{red}{[t_1.k = t_2.k] \times [t_1.a \geq 12]} \times & \\
 & \qquad \; \Big(\sum_{t_3} [t_3.k = t_1.k ] \times [t_3.a = t_1.a] \times R(t_3)\Big) \times R(t_2) & \text{\rref{prod-comm}} \\
 \quad & \eq  \sum_{t_1,t_2} [t_2 = t] \times [t_1.k = t_2.k] \times ([t_1.a \geq 12] \times & \\
 &  \qquad \; \textcolor{red}{\sum_{t_3} [t_3.k = t_1.k] \times[t_3.a = t_1.a] \times R(t_3) \times R(t_2)} & \text{\rref{pull-sum-r}} \\
 \quad & \eq \sum_{t_1,t_2, \textcolor{red}{t_3}} [t_2 = t] \times [t_1.k = t_2.k] \times [t_1.a \geq 12] \times & \\
 & \qquad \; [t_3.k = t_1.k] \times[t_3.a = t_1.a] \times R(t_3) \times R(t_2) & \text{\rref{pull-sum-l}}
\end{flalign*}
\end{small}
\vspace{-0.3in}
\caption{A SQL query $q$ (the second query shown in
  \figref{index-rewrite}), its semantics $\denote{q}$ in \ourring and its
  rewriting into sum-product normal form.}
\figlabel{index-nf}
\vspace{-0.1in}
\end{figure}

\setcounter{equation}{0}

To facilitate the automated equivalence proof, our algorithm,
\algname, first converts every {\newexp} into the \emph{Sum-Product
  Normal Form ({\firstnf})}.  Importantly, this conversion is done by
repeated applications of the \ourring axioms; thus, our system can
{\em prove} that any expression translated into \firstnf is
semantically equivalent to the original input \newexp.


%

\begin{definition}
\label{def:nf}
  A {\newexp} expression $E$ is in {\firstnf} if it has the following form:
\begin{itemize}
\item $E ::= T_1 + \ldots + T_n$
\item Furthermore, each term $T_i$ has the form:
  \begin{align*}
&  \sum_{t_1, \ldots, t_m} [b_1] \times \ldots \times [b_k] \times  \merely{E_s} \times \unot{E_n} \times M_1 \times \ldots \times M_j
  \end{align*}
  where each tuple variable $t_i$ ranges over $\Tuple({\schema}_i)$.
  Each expression inside the summation is called a {\em factor}.
  Multiple summations are combined into a single sum using
  Eq.~\eqref{eq:sum-symm}.  There is no summation in the case where $m=0$.
%
\item Each factor $[b_i]$ is a predicate.
\item There is exactly one factor $\merely{E_s}$, where $E_s$ is an
  expression in {\firstnf}.  When $E_s=\one$,  $\merely{E_s}$ can be omitted.
\item There is exactly one factor $\unot{E_n}$, where $E_n$ is an
  expression in {\firstnf}.  When $E_n = \zero$, $\unot{E_n}$ can be omitted.
\item Each factor $M_i$ is an expression of the form $R(t)$, for some
  relation name $R$, and some tuple variable $t$.
\end{itemize}
\end{definition}

\begin{theorem}\label{th:1strs}
  For any {\newexp} $E$, there exists an {\firstnf} expression $E'$
  such that $E = E'$ in any \ourring.
\end{theorem}

\begin{proof} We briefly sketch the proof idea here.  Formally,
  any {\newexp} $E$ can be rewritten into $E'$ in {\firstnf} using
  the following rewrite system:
%
\begin{align}
E_1 \times (E_2 + E_3 ) ~\rewrite~ & E_1 \times E_2 + E_1 \times E_3  \rlabel{distr-prod-plus-r}\\
(E_1 + E_2) \times E_3 ~\rewrite~ & E_1 \times E_3 + E_2 \times E_3 \rlabel{distr-prod-plus-l}\\
E_1 \times (E_2 \times E_3) ~\rewrite~ & (E_1 \times E_2) \times E_3  \rlabel{prod-assoc}\\
 \ldots \times M \times [b] \times \ldots \rewrite & \ldots \times [b] \times M \times \ldots \rlabel{prod-comm}\\
\sum_{t}\big(f_1(t) + f_2(t)\big) ~\rewrite~ & \sum_{t}f_1(t) + \sum_{t}f_2(t) \rlabel{distr-sum}\\
E \times \sum_{t} f(t) ~\rewrite~ & \sum_{t} E \times f(t) \rlabel{pull-sum-l}\\
\big(\sum_{t} f(t)\big) \times E ~\rewrite~ & \sum_{t} f(t) \times E \rlabel{pull-sum-r}\\
\merely{E_1} \times \merely{E_2} ~\rewrite~ & \merely{E_1 \times E_2} \rlabel{pull-merely} \\
\unot{E_1} \times \unot{E_2} ~\rewrite~ & \unot{E_1 + E_2} \rlabel{pull-not}
\end{align}

Each rule above corresponds to an axiom of {\ourring}s.
Rule~(\ref{rule:distr-prod-plus-r})-(\ref{rule:prod-comm}) are axioms
of commutative semirings, while the rest are axioms of {\ourring}s.  All
rewrite rules in \firstnf are unidirectional and guaranteed to
make progress. For example Rule~(\ref{rule:distr-prod-plus-r}) and
(\ref{rule:distr-sum}) apply distributivity to remove any $+$
inside each $T_i$.  Rule~(\ref{rule:prod-assoc}) and
(\ref{rule:prod-comm}) normalize the $\times$ expressions so they remain
 left associative with all boolean expressions
 moved to the left.
 The last two rules, Rule (\ref{rule:pull-merely}) and (\ref{rule:pull-not}), consolidate a product of multiple
factors into a single factor.
\end{proof}

Figure~\ref{fig:index-nf} shows how a \newexp is converted into \firstnf by applying Rules~(\ref{rule:prod-comm}, \ref{rule:pull-sum-r}), and (\ref{rule:pull-sum-l}). In general, the rules
described above are applied recursively until none of them is applicable.



\section{Integrity Constraints}

\seclabel{constraints}

Our system checks the equivalence of two SQL queries in the presence
of a set of integrity constraints: keys, foreign keys, views, and
indexes (see \figref{sql-syntax}).  In this section we introduce a new
axiomatic interpretation of integrity constraints using identities in
a {\ourring}.


\subsection{Axiomatic Interpretation of Constraints}
\seclabel{int-constraints}




\paragraph{Key Constraints.} To the best of our knowledge, key
constraints have not to date been defined for semiring semantics.
Therefore, we give the following definition.
\begin{definition} \label{def:key} Let $R$ be a relation with schema
  $\sigma$, and let $k$ be an attribute (or a set of attributes).  The
  \texttt{KEY} constraint is the following identity, for all
  $t, t' \in \Tuple({\schema})$:
\begin{align*}
  [\tuple.k = \tuple'.k] \times  R(\tuple) \times R(\tuple') ~~\eq~~ [\tuple = \tuple'] \times R(\tuple)
\end{align*}
%
\end{definition}

For each key constraint in the SQL specification the system generates
one such identity,
adds it as an axiom, and uses it later to prove
equivalences of SQL expressions.
We show two simple properties implied
by the key constraint.  First, setting $t=t'$, we have
$(R(t))^2 = R(t)$ for every tuple $t$.  Second, for two tuples
$t \neq t'$ such that $t.k = t'.k$, we have
$R(t)\times R(t')= \zero$. Therefore, if the \ourring\ is the
semiring of natural numbers, $\N$, then $R.k$ is a standard key:
the first property implies that $R(t) \in \set{0,1}$, and the second
implies that $R(t)=0$ or $R(t')=0$. Hence, only one tuple with
a given key may occur in $R$, with its multiplicity being 1:

\begin{theorem}
  If $R.k$ satisfies the key constraint (Def.~\ref{def:key}) over the
  \ourring\ of natural numbers $\N$, then $R.k$ is a standard key.
\end{theorem}

We briefly discuss our choice in Def.~\ref{def:key}. The standard
axiom for a key is
$\forall t, t' . (t \in R \wedge t' \in R \wedge t.k=t'.k
~\Rightarrow~ t=t')$~\cite{AbiteboulHV95}.
However, it applies only to set semantics and uses a first-order
sentence instead of an identity in a semiring. An alternative
attempt at defining a key is:
\begin{align}
\sum_{\tuple} R(\tuple) \times [\tuple.k=e] \leq \one \label{eq:key2}
\end{align}
This says that the sum of all multiplicities of tuples that have their keys
equal to a constant $e$ must be $\leq \one$.  However, (\ref{eq:key2})
requires an ordered semiring, which leads to additional complexity, as
argued earlier.  In contrast, Def.~\ref{def:key} does not require
order and is a simple identity, i.e., it has the form $E_1 = E_2$.
%
A better attempt is to state:
\begin{align}
  R(\tuple') \times \sum_{\tuple} R(\tuple) \times [\tuple.k=e] ~~=~~ R(t')\label{eq:key3}
\end{align}
This axiom is an identity and, furthermore, it can be shown that
Eq.(\ref{eq:key2}) implies Eq.(\ref{eq:key3}) in
any ordered semiring; in
other words, Eq.(\ref{eq:key3}) seems to be the right reformulation of
Eq.(\ref{eq:key2}) without requiring order.  On the other hand,
Eq.(\ref{eq:key3}) already follows from our key identity
(Def.~\ref{def:key}): simply sum both sides over $t$ and observe that the
RHS is equal to $R(t')$.  We prove here a consequence of the key
constraint, which we use in \secref{alg-example} to prove
some non-trivial SQL identities described by~\cite{PiraheshHH92}:

\begin{theorem}
\label{th:key-merely}
If $R.k$ satisfies the key constraint (Def.~\ref{def:key}), then the following {\newexp} is preserved under the squash operator $\merely{\cdot}$,
for any \newexp\  $E$,  expression $e$,  predicate $b$:
\begin{align*}
\sum_{t}[b] \times & \merely{E} \times [t.k = e] \times R(t)  = \nonumber \\
& \merely{\sum_{t}[b] \times \merely{E} \times [t.k = e] \times R(t)}
\end{align*}
\end{theorem}

\begin{proof}
  By Eq.\eqref{eq:square-merely} in \secref{ring}, it suffices to show that the LHS
  squared equals itself.  From Def.~\ref{def:key}, we
  derive:\footnote{We also used $[b]^2 = [b]$ and
    $\merely{E}^2 = \merely{E}$.}
\begin{align*}
[t.k = e] \times [t'.k = e] \times [b]^2 \times \merely{E}^2 \times R(t) \times R(t') & =\\
[t = t'] \times [t.k = e] \times [b] \times \merely{E} \times R(t) \nonumber
\end{align*}
Next, we sum over $t$ and $t'$ on both sides:
\begin{align*}
\sum_{t,t'}[t.k = e] \times [t'.k = e] \times [b]^2 \times \merely{E}^2 \times R(t) \times R(t') & =  \\
\sum_{t,t'}[t = t'] \times [t.k = e] \times [b] \times \merely{E} \times R(t) &
\end{align*}
By applying Eq.~\eqref{eq:prod-sum-assoc} in \secref{ring} twice on the left, and
Eq.~\eqref{eq:sum-subst-eq} in \secref{denotation} on the right:
\begin{small}
\begin{align*}
& \Big(\sum_{t} [b] \times \merely{E} \times [t.k = e] \times R(t)\Big) \times \
\Big(\sum_{t'} [b] \times \merely{E} \times [t'.k = e] \times R(t')\Big) \\
& = \sum_{t}[b] \times \merely{E} \times [t.k = e] \times R(t)
\end{align*}
\end{small}
Hence Theorem~\ref{th:key-merely} follows from  Eq.~\eqref{eq:square-merely} in \secref{ring}:
$x^2=x$ implies $\merely{x}=x$.
\end{proof}


\paragraph{Foreign Key Constraints.} We now define foreign keys in a
\ourring:
%
%
\begin{definition} \label{def:fk} Let $S, R$ be two relations, and
  $k',k$ be two attributes (or two sets of attributes),
  in $S$ and $R$, respectively.
  The \texttt{FOREIGN KEY} constraint from $S.k'$ to
  $R.k$ is the following:
\begin{align*}
S(\tuple') = &\; S(\tuple') \times \sum_{\tuple}  R(\tuple) \times [\tuple.k = \tuple'.k']
\end{align*}
%
\end{definition}

If $S(\tuple')\neq 0$ and the \ourring\ is the standard semiring $\N$,
then this definition implies
$\sum_{\tuple} R(\tuple) \times [\tuple.k = \tuple'.k']=1$, i.e., a tuple in $R$ has the same value $k$, the tuple
is unique, and its multiplicity is 1. There is no constraint on the
multiplicity of $S(\tuple')$. Thus:

\begin{theorem}
  If $S.k', R.k$ satisfies the foreign key constraint (Def.~\ref{def:fk}) over
  the \ourring\ of natural numbers $\N$, then $R.k$ is a standard key
  in $R$, and $S.k'$ is a standard foreign key to $R.k$.
\end{theorem}

\paragraph{Views and Indexes.} We convert a view definition in \figref{sql-syntax}
 into an assertion $v(\tuple) = q(\tuple)$; every
occurrence of $v$ in a query is inlined according to its definition.
We follow the GMAP approach~\cite{TsatalosSI96} and
consider an index to be a view definition that consists of the
projection on the key and the attribute to be indexed.
%
%
%
For example, an index $I$ on the attribute $R.a$ is expressed as:
\[ I := \texttt{SELECT $x.a$, $x.k$ FROM $R$ $x$}\]
where $k$ is the \texttt{KEY} (Def.~\ref{def:key}) of R.  The indexes are treated as views
and inlined in the main query when compiling to {\newexp}s.

\subsection{SQL Equivalence with Integrity Constraints}
\seclabel{ex-constraints}

We can now formally define the equivalence of two SQL expressions
under \ourring semantics.
\begin{definition} \label{def:u:equivalence}
Consider two queries, $q_1$ and $q_2$, along with definitions of
schemas, base relations, and constraints (Fig.~\ref{fig:sql-syntax}).
$q_1$ and $q_2$ are U-equivalent if, for any \ourring and
interpretation of the base relations, 
when all key and foreign key constraints are satisfied, then the following identity holds:
$\denote{q_1} = \denote{q_2}$.\footnote{We assume all views are inlined into $q_1$ and $q_2$ as discussed above.}
\end{definition}
We will refer to U-equivalence simply as \emph{equivalence} when the context is clear. We illustrate this concept with an example.


\begin{example}
  \label{ex-index}
  We show how to formally prove that the optimization rule at the top
  of~\figref{index-rewrite} is correct by showing that
  $Q_1$ is equivalent to its rewrite, $Q_2$ using an index lookup.
 Recall that $R$ has key $k$, and $I$ is an index on $R.a$.

%


The {\newexp} of $Q_1$ is:
\[ Q_1(t) = R(t) \times [t.a \geq 12] \]
After inlining the definition of $I$, $Q_2$ is the query shown
in Fig.\ref{fig:index-nf}, and its {\newexp} in {\firstnf} is:
  \begin{flalign*}
    Q_2(t) & \eq \sum_{t_1,t_2,t_3} [t_2 = t] \times [t_1.k = t_2.k] \times [t_1.a \geq 12] \times  \\
 & \qquad \; [t_3.k = t_1.k] \times[t_3.a = t_1.a] \times R(t_3) \times R(t_2)
  \end{flalign*}
  Since $t_1$ is a tuple returned by $I$, its schema consists of the
  two attributes $a$ and $k$; hence, $[t_3.k = t_1.k]$ and
  $[t_3.a = t_1.a]$ imply $[t_1 = (t_3.k,t_3.a)]$. We use
  Eq.~\eqref{eq:sum-subst-eq} to remove the summation over $t_1$,
  the two equalities $[t_3.k = t_1.k]$ and $[t_3.a = t_1.a]$, and
  substitute $t_1.k$ with $t_3.k$ and $t_1.a$ with $t_3.a$ to get:
  \begin{flalign*}
    Q_2(t) & \eq \sum_{t_2,t_3} [t_2 = t] \times [\red{t_3.k} = t_2.k] \times [\red{t_3.a} \geq 12] \times R(t_3) \times R(t_2)
  \end{flalign*}
Applying the key constraint definition (Def.~\ref{def:key}) on $Q_2(t)$, we get:
\[
 Q_2(t) \eq \sum_{t_2,t_3} [t_2 = t] \times [t_3.a \geq 12] \times R(t_2) \times [t_2 = t_3]
\]
Now, we use Eq.~\eqref{eq:sum-subst-eq} to remove the summation over
$t_3$ as $t_2 = t_3$:
\[
 Q_2(t) \eq \sum_{t_2} [t_2 = t] \times [\red{t_2.a} \geq 12] \times R(t_2)
\]
And similarly remove the summation over $t_2$ since $t_2 = t$:
\[
 Q_2(t) \eq  [\red{t.a} \geq 12] \times R(\red{t})
\]
This shows that $Q_1$ and $Q_2$ are equivalent by Def.~\ref{def:u:equivalence}.
\end{example}

The example above reveals a strategy for proving queries: first, express both queries in
{\firstnf}. Then, repeatedly reduce the expression using the available constraints until
the they become isomorphic and hence are equivalent.
We formally present our equivalence deciding algorithm in the next section.

{\bf Discussion.} Definition~\ref{def:u:equivalence} is sound because
two U-equivalent queries are also equivalent under the standard
interpretation in the semiring of natural numbers.  However, the
definition is not complete: there exists SQL queries that are
equivalent under the standard semantics but not U-equivalent.  One
such example consists of queries that are equivalent over all finite
relations but not equivalent over infinite relations.  For example,
there exists sentences $\varphi$ in First Order Logic, called {\em
  infinity axioms}, that are always false when $R$ is finite, but can
be satisfied by an infinite $R$; for example, $\varphi$ may say ``$R$
is non-empty; for every $x$ occurring in $R$ there exists a unique $y$
such that $R(x,y)$ (i.e. $R$ is a function); the mapping $x\mapsto y$
is injective; and it is not surjective.''\footnote{A different, shorter
  infinity axiom is given in
  \cite[pp.307]{DBLP:books/sp/BorgerGG1997}:
  $\varphi \equiv \forall x \exists y \forall z(\neg R(x,x) \wedge
  R(x,y) \wedge (R(y,z) \rightarrow R(x,z)))$.}
Such a sentence $\varphi$ can be converted in SQL query $Q$ that
returns $\emptyset$ when $\varphi$ is false, and returns $1$ when
$\varphi$ is true.  Then, on any finite database $Q$ is equivalent to
the empty-set query $Q'$ (e.g, written in SQL as
$\texttt{select distinct 1 from R where } 0 \neq 0$), but $Q,Q'$ are
not U-equivalent, because they are distinct over the \ourring\
$\bar \N$.

%


\section{Decision Procedure for SQL}
\seclabel{alg}

We now present \algname (\newexp Decision Procedure) for checking the equivalence of two \newexps with constraints,
where {\em equivalence} is the U-equivalence given by Definition~\ref{def:u:equivalence}.
\reviewerthree{As we are unaware of any theorem provers that can automatically reason about 
the equivalences of \newexps, we implement our decision procedure using proof assistants, which 
ensures that our procedure implementation is {\em provably correct} given the \newexp axioms that we have defined.}


\algname supports the SQL fragment in~\figref{sql-syntax}.
The input SQL queries are evaluated under mixed set and bag semantics,
which is the semantics that most real-world database systems use.
We show that \algname is sound and also complete when the input are
Union of Conjunctive Queries (UCQ) and evaluated under set semantics
only,  or under bag semantics only.
To the best of our knowledge, this is the first
algorithm implemented that can check the equivalence of UCQ with integrity constraints and evaluated under set or bag semantics.

At a high level, \algname proceeds recursively on the structure of a
\newexp\, as shown in Alg.~\ref{alg:main}.  Recall from
Def.~\ref{def:nf} the structure of a normalized {\newexp} $E$ and of a
term $T$:
%
%
%

\begin{small}
\begin{align*}
E ::= & T_1 + \ldots + T_n \\
T_i ::= & \sum_{t_1, \ldots, t_m} [b_1] \times \ldots \times [b_k] \times  \merely{E'} \times \unot{E''} \times M_1 \times \ldots \times M_j
\end{align*}
\end{small}

%
\algname takes two \newexps ($E_1$ and $E_2$),
and a set of  integrity constraints ($C$) in the form of \newexp identities. It first calls \texttt{canonize} (line 2) to transform each \newexp\ into a canonical representation (to be discussed \secref{2nf}).
Then, \algname proceeds recursively on the structure of the two
canonical \newexps by calling \termalgname (line 7) shown in Alg.~\ref{alg:pterm} to check the equivalence of each term. 
Similarly, procedure \termalgname calls \setalgname (line 4)
shown in Alg.~\ref{alg:merely} to check the equivalence of two squashed \newexps
(\newexps in the form of $\merely{E}$). We discuss the details below.



\subsection{Canonical Form}
\seclabel{2nf}

As we have illustrated in Ex~\ref{ex-index}, to prove the equivalence of SQL queries under integrity constraints, we rewrite the input query \newexps using the axiomatic interpretations of these constraints as shown in \secref{int-constraints}.
We call these rewritten \newexps the canonical form of \newexps under integrity constraints.

\begin{algorithm}
\caption{Canonization}\label{alg:tonf}
\begin{algorithmic}[1]
\Procedure{\texttt{canonize}}{$E, C$}
\Statex \Comment{$E$ is an \newexp in \firstnf; $C$ is a set of constraints}
\State $E' \gets \texttt{TC}(E)$                   \hspace{0.2in}\Comment{Transitive closure of equalities}
\State $E' \gets $ \emph{recursively} and \emph{repeatedly} apply Eq.~\eqref{eq:sum-subst-eq} on $E'$
\For{$c \in C$}
   \State $E' \gets$ \emph{recursively} and \emph{repeatedly} apply Def.~\ref{def:key} and
Theorem~\ref{th:key-merely} on $E'$
   \State $E' \gets$ \emph{recursively} and \emph{repeatedly} apply Def.~\ref{def:fk} on $E'$
\EndFor
\State \textbf{return} $E'$ \hspace{0.2in}\Comment{$E'$ is now in canonical form}
\EndProcedure
\end{algorithmic}
\end{algorithm}

Algorithm~\ref{alg:tonf} shows the detail of canonization.
To convert a \newexp $E$ from \firstnf to canonical form,
we rewrite $E$ by the axioms and definitions discussed in \secref{axioms} and \secref{constraints}.
The algorithm contains the following four rewrites:

\begin{enumerate}
\item Apply the transitivity of the equality predicate (line 2): $[e_1 = e_2]
  \times [e_2 = e_3] ~~=~~ [e_1=e_2] \times [e_2 = e_3] \times
  [e_1 = e_3]$.
\item Remove unnecessary summations using Eq.~\eqref{eq:sum-subst-eq}
(line 3).
\item For each key constraint, rewrite $E$ using the identities defined
in Def.~\ref{def:key} and Theorem~\ref{th:key-merely} (line 5).
\item For each foreign key constraint, rewrite $E$ using the identity defined in Def.~\ref{def:fk} (line 6).
\end{enumerate}

When using an identity to rewrite $E$,
we find a matching subexpression on $E$ with the identity's LHS,
and then replace it with the RHS.
More importantly, these rewrites are performed
\emph{recursively} on the structure of $E$,
and \emph{repeatedly} until the following termination conditions:
1) The first $3$ rewrites (line 2, 3, and 5) terminate until no further rewrite can
be applied. 2) When applied on squashed expressions,
e.g., $\merely{E'}$, the last rewrite (line 6) is repeatedly applied
until the rewritten squashed expression is equivalent
 to the one before applying the rewrite.
The equivalence of the squashed expressions (before and after applying the rewrite)
is checked using procedure \setalgname that will be presented
in Algorithm~\ref{alg:merely} (\secref{eq-alg}).
3) When applied on other expressions,
the last rewrite (line 6) is repeatedly applied until no new base relation is introduced.

\paragraph{Does \texttt{canonize} terminate?} The rewrites using key and
foreign key definitions resemble the chase procedure~\cite{PopaDST00}.
The difference is that our rewrites are on \newexps rather than on relational queries.
\reviewertwo{It is known that the chase procedure may not
  terminate~\cite{FaginKMP03ICDT}, and so may our rewrite steps; for
  example, a cycle in the key/foreign-key graph may lead to
  non-termination.  While this is possible in theory, we did not
  encounter any case that does not terminate, as our evaluation 
  in Sec.~\ref{sec:evaluation} shows.}

\subsection{Equivalence of {\newexps}}
\seclabel{eq-alg}

We now present the algorithm for checking if two \newexps are equivalent after converting them into canonical form.

\begin{algorithm}
\caption{{\algname}: \newexp Decision Procedure}\label{alg:main}
\begin{algorithmic}[1]
\Procedure{\algname}{$E_1, E_2, C$}  \hspace{0.2in}\Comment{$E_1$ and $E_2$ are in {\firstnf}}
  \State $E_1 \gets \texttt{canonize}(E_1, C), E_2 \gets \texttt{canonize}(E_2, C)$
  \Statex \hskip\algorithmicindent\Comment{$E_1$ is in the form of  $T_{1,1} + \ldots + T_{1,n}$, and}
  \Statex \hskip\algorithmicindent\Comment{$E_2$ is in the form of  $T_{2,1} + \ldots + T_{2,m}$}
  \If{$m \neq n$}
  \State \textbf{return false}
  \EndIf
  \For{$p \in \mathcal{P}([T_{1,1}, \ldots, T_{1,n}])$}
  \If{$\forall i \in \{1,\ldots, n\}, \termalgname(p[i], T_{2,i}, C) ==$ \textbf{true}}
          \State \textbf{return true}
      \EndIf
  \EndFor
  \State \textbf{return false}
\EndProcedure
\end{algorithmic}
\end{algorithm}

Alg.~\ref{alg:main} shows the detail of \algname for checking the equivalence of two \newexps.
On line 2, \algname first converts the two input \newexps ($E_1$ and $E_2$) to
canonical forms under integrity constraints as described in~\secref{2nf}.  Recall that a \newexp\ is a sum of terms $E = T_1 + \ldots + T_n$.  To check for U-equivalence of $E_1$ and $E_2$, \algname searches for an isomorphism between each of their constituent terms $T_i$ in line 3-10.
It returns false if the number of terms of $E_1$ and $E_2$ are not
equal (line 4). Otherwise, it searches for a permutation $p$ of $E_1$'s
terms (here $\mathcal{P}$ represents the set of all possible permutations of its arguments) such that each pair of terms in $p$ and $E_2$ are equivalent. This is determined by calling \termalgname in line 8, which we discuss next.

\paragraph{Equivalence of Terms}
\vspace{1mm}

The \termalgname procedure shown in Alg.~\ref{alg:pterm} checks the equivalence between terms. Recall that a term is an unbounded summation of the form $T = \sum_{t_1, \ldots, t_m}(\ldots)$. To check for the equivalence of the following input terms:
\begin{small}
\begin{align*}
 T_1 = & \sum_{\vv{t_1}} [b_{1,1}] \times \ldots \times [b_{1,k}] \times  \merely{E_{1,1}} \times \unot{E_{1,2}} \times  M_{1,1} \times \ldots \times M_{1,j} \\
T_2 = & \sum_{\vv{t_2}} [b_{2,1}] \times \ldots \times [b_{2,m}] \times  \merely{E_{2,1}} \times \unot{E_{2,2}} \times M_{2,1} \times \ldots \times M_{2,\ell}
\end{align*}
\end{small}
\termalgname searches for an isomorphism between $T_1$ and $T_2$. 
Let $h \in \mathcal{BI}_{\vv{t_2}, \vv{t_1}}$
be a bijection from the set of variables that $T_2$ sums over ($\vv{t_2}$) to
the set of variables that $T_1$ sums over ($\vv{t_1}$) (line 2).\footnote{\small For ease of presentation, we use the vectorized notation $\vv{t}$
as a shorthand for $t_1, t_2, \ldots$}
\termalgname find an isomorphism if, after substituting $\vv{t_1}$
with $p(\vv{t_2})$ in $T_2$ (we call this new expression $T_2'$) (line 3),
the equivalence of $T_1$ and $T_2'$ is checked as following (line 4):

\begin{itemize}
\item The predicate parts of two terms are equivalent:
  $[b_{1,1}] \times \ldots \times [b_{1,k}] ~ = ~ [b_{2,1}] \times
  \ldots \times [b_{2,m}]$.  This requires checking that two Boolean
  expressions are equivalent.
We check the equivalences of boolean expressions using the congruence procedure
\cite{NelsonO80JACM}, which first computes the equivalent classes of variables and
function applications and then checks for equivalence of the expressions using the equivalent classes.
For example, $[a = b] \times [c = d] \times [b = e] \times [f(a) = g(d)]$
is equivalent to $[a = b] \times [a = e] \times [c = d] \times [f(e) = g(c)]$,
since there are the following equivalent classes:
\[ \{a, b, e\}, \; \{c, d\},\; \{f(a), f(e)\}, \{g(c), g(d)\} \]
A predicate in $T_1$ (e.g., $[b_{1,i}]$) may contains aggregate functions (for example,
  $[\texttt{avg}(\cdots) = t.a]$). These aggregate functions are treated as uninterpreted functions like $f$ and $g$ in the example above.

\item $\merely{E_{1,1}}$ is U-equivalent to $\merely{E_{2,1}}$. We explain
  the procedure for checking equivalences of squashed \newexps,
  \setalgname, in the next section. 

\item The negated expressions $\unot{E_{1,2}}$ and $\unot{E_{2,2}}$ are
  U-equivalent.  This is checked by calling \algname recursively in line 4.

\item The two terms have the same number of $M$ subterms, i.e., $j = \ell$, and the terms $M_{1,1}, \ldots, M_{1,j}$ are identical to the terms $M_{2,1}, \ldots, M_{2,\ell}$. Recall that each $M_{i_1, i_2}$ has the form $R(t)$ for some relation name $R$ and some tuple variable $t$ (Def.~\ref{def:nf}).
\end{itemize}


\begin{algorithm}
\caption{{\tt TDP}: Decision Procedure for Terms}
\label{alg:pterm}
\begin{algorithmic}[1]
    \Procedure{\termalgname}{$T_1, T_2, C$}
  \Statex \Comment{$T_1, T_2$ is in \firstnf, $C$ is a set of constraints. In particular,}

  \Statex \Comment{$T_1$ has the form $\sum_{\vv{t_1}} [b_{1,1}] \times \ldots \times [b_{1,k}] \times  \merely{E_{1,1}} \times$}
 \Statex \Comment{\hspace{1in}$\unot{E_{1,2}} \times M_{1,1} \times \ldots \times M_{1,j}$, and}

  \Statex \Comment{$T_2$ has the form $\sum_{\vv{t_2}} [b_{2,1}] \times \ldots \times [b_{2,m}] \times  \merely{E_{2,1}} \times$}
  \Statex \Comment{\hspace{1in}$\unot{E_{2,2}} \times M_{2,1} \times \ldots \times M_{2,\ell}$}

  \For{$p \in \mathcal{BI}_{\vv{t_2}, \vv{t_1}}$}
     \State $T_2' \gets p(T_2)$   \hspace{0.2in}\Comment{substitute $\vv{t_2}$ in $T_2$ using $p(\vv{t_2})$}
     \If{$T_1 == T_2'$}
       \State \textbf{return true}
     \EndIf
  \EndFor
  \State \textbf{return false}
\EndProcedure
\end{algorithmic}
\end{algorithm}
\paragraph{Equivalences of  Squashed  Expressions}
\vspace{1mm}

Finally, Alg.~\ref{alg:merely} shows \setalgname, the procedure for checking the equivalence of two squashed
\newexps, $\merely{E_1}$ and $\merely{E_2}$.
Recall that squash expressions $\merely{\cdot}$ are used to model the semantics of (sub-)queries with the \texttt{DISTINCT} operator.

The first step of \setalgname is to remove any nested squash subexpressions (line 2).
We do so by applying the following lemma:
%
\begin{lemma} \label{lem:reduce-merely}  The following holds in any \ourring:
\begin{align} \label{eq:reduce-merely}
  \merely{a \times \merely{x} + y } \eq \merely{a\times x + y}
\end{align}
\end{lemma}
\begin{proof}
  By Eq.~\eqref{eq:merely-plus} in \secref{axioms}, $LHS = \merely{\merely{a \times \merely{x}} + y }$. And by Eq.~\eqref{eq:merely-times}, $LHS = \merely{\merely{a} \times \merely{\merely{x}} + y }$. By Eq.~\eqref{eq:merely-plus} (when $y=\zero$ in Eq.~\eqref{eq:merely-plus}), $\merely{\merely{x}} = x$; thus:
\[ LHS = \merely{\merely{a} \times \merely{x} + y } \]
  Apply Eq.~\eqref{eq:merely-times} again:
\[ LHS = \merely{\merely{a \times x} + y } \]
  And apply Eq.~\eqref{eq:merely-plus} from right to left:
\[ LHS = \merely{a \times x + y } \]
\end{proof}

\begin{algorithm}
\caption{{\setalgname}: Decision Procedure for Squashed Expressions}\label{alg:merely}
\begin{algorithmic}[1]
\Procedure{{\setalgname}}{$\merely{E_1}, \merely{E_2}, C$} \hspace{0.2in}\Comment{$E, E'$ are in {\firstnf}}
   \State remove $\merely{}$ inside $E_1, E_2$ by
   applying Lem.~\ref{lem:reduce-merely}
   \State $\merely{E_1} \gets \merely{\texttt{canonize}(E_1, C)}$
   \Statex \hskip\algorithmicindent \Comment{$\merely{E_1}$ is now in the form of $\merely{T_1 + \ldots + T_m}$}
   \State{$\merely{E_2} \gets \merely{\texttt{canonize}(E_2, C)}$}
   \Statex \hskip\algorithmicindent \Comment{$\merely{E_2}$ is now in the form of $\merely{T_1' + \ldots + T_n'}$}
   \State{\textbf{return} $\forall i \, \exists \; j . \; \texttt{minimize}(\merely{T_i}) == \texttt{minimize}(\merely{T_j'})$}
   \Statex{\hskip\algorithmicindent  $ \quad \&\& \; \forall j \, \exists \; i . \; \texttt{minimize}(\merely{T_j'}) == \texttt{minimize}(\merely{T_i})$}
\EndProcedure
\end{algorithmic}
\end{algorithm}

\setalgname then converts the expressions
$\merely{E_1}$ and $\merely{E_2}$ to canonical forms under constraints
by calling \texttt{canonize} on (lines 3-4).
After that, \setalgname checks the equivalence of two
expressions $\merely{T_1 + \cdots + T_m}$ and
$\merely{T_1' + \cdots + T_m'}$.

If $T_i$ and $T_i'$ (for $i = 1, \ldots, m$) are conjunctive queries,
then $\merely{T_1 + \cdots + T_m}$ and
$\merely{T_1' + \cdots + T_m'}$ represent two queries in the class of unions
of conjunctive queries under set semantics.
A classical algorithm exists for checking the equivalence of such \\ 
queries~\cite{SagivY80JACM}:
given $Q = q_1 \vee \cdots \vee q_m$ and $Q' = q_1' \vee \cdots \vee q_n'$, where each $q$ is a conjunctive query,
equivalence is established by checking whether $Q \subseteq Q'$ and $Q' \subseteq Q$.  The former is checked by showing that for every $i$, there exists a $j$ such that $q_i \subseteq q_j$ which, in turn, requires checking
for a homomorphism from $q_j$ to $q_i$.
$Q' \subseteq Q$ is checked similarly.

\reviewerthree{
However, it is challenging to express the homomorphism checking algorithm in~\cite{SagivY80JACM} purely using \ourring axioms since \ourring axioms are all identities (equations). 
There is no notion of inclusion or ordering in \ourring.
To check the equivalences of two canonized squashed expressions, $\merely{E_1}$ and $\merely{E_2}$,
we instead minimize each term, $T_i$, inside $\merely{E_1}$ 
and $\merely{E_2}$ using only the \ourring axioms (this is implemented in the \texttt{minimize} procedure) and then check the
syntactic equivalences of the minimized terms. This procedure is equivalent to 
minimizing conjunctive queries~\cite{AbiteboulHV95}. Thus, our equivalence checking procedure is sound and complete for squashed expressions derived from UCQs.
For squashed expressions that represent SQL queries beyond UCQs, \setalgname is not complete, however, our procedure is still sound.}
\reviewerthree{
Next, we explain how to minimize a term in a squashed expression using only \ourring axioms.
\texttt{minimize} follows exactly the same steps as shown in the example, except needs to be
performed on all possible pairs of summation variables.} 

\begin{example}
  \label{ex:homo}

We illustrate \setalgname on  these two queries:
\begin{lstlisting}[style=sql]
SELECT DISTINCT x.a FROM R x, R y   -- Q1
SELECT DISTINCT R.a FROM R          -- Q2
\end{lstlisting}

It first converts $Q_1$ to a {\newexp}s canonical form (line 2-4):
\begin{align*}
   Q_1(t) \eq & \merely{ \sum_{t_1, t_2} [t_1.a = t] \times R(t_1) \times R(t_2)} \\
   \stackrel{\mbox{Eq.(\eqref{eq:lem})}}{=} & \merely{ \sum_{t_1, t_2} ([t_1 = t_2] + [t_1 \neq t_2])  \times [t_1.a = t] \times R(t_1) \times R(t_2)}
\end{align*}
%
%
%
Next, it uses the rewrite rules in \secref{fnf} to convert it to \firstnf:
\begin{align*}
Q_1(t) = & \parallel \sum_{t_1, t_2} [t_1 = t_2] \times [t_1.a = t] \times R(t_1) \times R(t_2) + \\
 & \quad \sum_{t_1, t_2} [t_1 \neq t_2] \times [t_1.a = t] \times R(t_1) \times R(t_2) \parallel
\end{align*}
\reviewerthree{   
Using Eq.~\eqref{eq:sum-subst-eq}, it removes the summation over
$t_2$ as $t_1 = t_2$:}
\reviewerthree{
\begin{align*}
Q_1(t) = & \parallel \sum_{t_1} [t_1.a = t] \times R(t_1) \times R(t_1) + \\
 & \quad \sum_{t_1, t_2} [t_1 \neq t_2] \times [t_1.a = t] \times R(t_1) \times R(t_2) \parallel
\end{align*}
By Eq.~\eqref{eq:squash-sum-squash} and Eq.~\eqref{eq:merely-square} ($\merely{R(t_1) \times R(t_1)} = \merely{R(t_1)}$), we have:
\begin{align*}
Q_1(t) = & \parallel \sum_{t_1}  [t_1.a = t] \times R(t_1) + \\
 & \quad \sum_{t_1, t_2} [t_1 \neq t_2] \times [t_1.a = t] \times R(t_1) \times R(t_2) \parallel
\end{align*}
Now, we can factorize the expression inside squash using Eq.~\eqref{eq:sum-distr}:
\begin{align*}
Q_1(t) = & \parallel \sum_{t_1}  [t_1.a = t] \times R(t_1) (1 + 
\sum_{t_2} [t_1 \neq t_2] \times R(t_2) \parallel
\end{align*}
Finally, we simplify $Q_1(t)$ using Eq.~\eqref{eq:merely-zero-one} and Eq.~\eqref{eq:merely-times}:
\begin{align*}
Q_1(t) = & \parallel \sum_{t_1}  [t_1.a = t] \times R(t_1) \times 1 \parallel \\
       = & \parallel \sum_{t_1}  [t_1.a = t] \times R(t_1) \parallel = Q_2(t)
\end{align*}
thus, proving that $Q_1(t)$ and $Q_2(t)$ are equivalent.}
\end{example}

\subsection{Soundness and Completeness}

We now show that UDP, our procedure for checking the equivalence of two \newexps, is sound. Furthermore, it is complete if two \newexps are
unions of conjunctive queries evaluated under set or bag semantics.

\begin{theorem}
  Algorithm~\ref{alg:main} is sound. For any pair of
  SQL queries, if algorithm~\ref{alg:main} returns \textbf{true}, then the pair is equivalent under the standard SQL semantics \cite{Date89AW}.
\end{theorem}

\begin{proof}
  All transformations in
  algorithms~\ref{alg:tonf} and~\ref{alg:main} are based on
  axioms of \ourring and proven identities.  Since the standard SQL semantics \cite{Date89AW} is based on the semiring of natural numbers $\N$, it follows that the equivalence also holds under the standard semantics.
\end{proof}

When there is no integrity constraint and $Q$ is a CQ evaluated under bag semantics, i.e., $Q$ has the form \texttt{SELECT p FROM R1 t1, ... , Rn tn WHERE b},
and \texttt{b} is a conjunction of equality predicates, then $\denote{Q}$ has a unique canonical form, namely
\[
\texttt{canonize}(\denote{Q}, \emptyset) = \sum_{t_1,...,t_n} [b] \times
R_1(t_1) \times \ldots \times R_n(t_n)
\]

In addition, if $Q$ is a CQ evaluated under set semantics, i.e.,
$Q$ has the form \texttt{SELECT DISTINCT p FROM R1 t1, ..., Rn tn WHERE b},
then $Q$ has a similar unique canonical form
(the same form as bag semantics CQ but within $\merely{\cdot}$).

This allows us to prove the following two theorems on the completeness of UDP.

\begin{theorem}
\label{theorem:bag-sound}
Algorithm~\ref{alg:main} is complete for checking the equivalence of
Unions of Conjunctive Queries (UCQ) evaluated under bag semantics.
\end{theorem}

\begin{proof}
  Two UCQ queries under bag semantics are equivalent if and only if they are isomorphic~\cite{SagivY80JACM} (see also
  \cite[Theorem~4.3, 4.4]{CohenNS99}), implying that our algorithm is
  complete in this case.
\end{proof}

\begin{theorem}
\label{theorem:set-sound}
Algorithm~\ref{alg:main} is complete for checking the equivalence of
Unions of Conjunctive Queries (UCQ) evaluated under set semantics.
\end{theorem}

\begin{proof}
  The \newexp\ of a conjunctive query evaluated under set semantics is
  $\merely{\sum_{\vec t} [b_1] \times \cdots \times [b_n] \times
    R_1(t_1) \times \cdots \times R_j(t_j)}$,
  where all predicates $b_i$ are equalities $[t.a_1 = t'.a_2]$.
  In this case, \algname simply checks for the existence of a homomorphism for the input queries, which has been shown to be complete~\cite{SagivY80JACM}.
\end{proof}

\subsection{An Illustration}
\seclabel{alg-example}

We demonstrate how {\algname} works using an example rewrite evaluated under mixed set-bag semantics from the Starburst optimizer~\cite{PiraheshHH92}:

\begin{lstlisting}[style=sql, basicstyle=\small\ttfamily]
-- Q1
SELECT ip.np, itm.type, itm.itemno
FROM (SELECT DISTINCT itp.itemno as itn,
                      itp.np as np
      FROM price price
      WHERE price.np > 1000) ip, itm itm
WHERE ip.itn = itm.itemno;

-- Q2
SELECT DISTINCT price.np, itm.type, itm.itemno
FROM price price, itm itm
WHERE price.np > 1000 AND
      itp.itemno = itm.itemno;
\end{lstlisting}
Here \texttt{itemno} is a key of \texttt{itm}.

Below is the \newexp representing $Q_1(t)$:
\begin{small}
\[
\begin{array}{llll}
 Q_1(t) & \eq & \sum_{t_1, t_2} [\attr{t_1}{np} = \attr{t}{np}] \times [\attr{t_2}{type} = \attr{t}{type}] \times & \\
  & &  [\attr{t_2}{itemno} = \attr{t}{itemno}] \times [\attr{t_1}{itn} = \attr{t_2}{itemno}] \times   & \\
  & &  \parallel \sum_{t'} [\attr{t'}{itemno} = \attr{t_1}{itn}] \times [\attr{t'}{np} = t_1.np] \times & \\
  & & \; [\attr{t'}{np} > 1000] \times \rel{price}(t') \parallel \times \rel{itm}(t_2)  & \\
\end{array}
\]
\end{small}

Next, $Q_1(t)$ is canonized by \texttt{canonize} (called by \algname in line 2).
As tuple $t_1$ is generated by the subquery in \texttt{Q1}, it has two attributes: \texttt{np} and \texttt{itn}. Because $[\attr{t_1}{np} = \attr{t}{np})]$ and $[\attr{t_1}{itn} = \attr{t_2}{itemno}]$, the summation on $t_1$ is removed after applying Eq.~\eqref{eq:sum-subst-eq} in line 3 in \texttt{canonize}:
\begin{small}
\[
\begin{array}{llll}
 Q_1(t) & \eq &  \sum_{t_2} [\attr{t_2}{type} = \attr{t}{type}] \times [\attr{t_2}{itemno} = \attr{t}{itemno}]  \times   & \\
  & &  \parallel \sum_{t'} [\attr{t'}{itemno} = \attr{t}{itemno}] \times [\attr{t'}{np} = t.np] \times & \\
  & & \; [\attr{t'}{np} > 1000] \times \rel{price}(t') \parallel \times \rel{itm}(t_2)  & \\
\end{array}
\]
\end{small}
Since \texttt{ite.itemno} is a key, Theorem~\ref{th:key-merely} is applied (line 5 in \texttt{canonize}):
\begin{small}
\[
\begin{array}{llll}
 Q_1(t) & \eq &  \parallel \sum_{t_2, t'} [\attr{t_2}{type} = \attr{t}{type}] \times [\attr{t_2}{itemno} = \attr{t}{itemno}]  \times   & \\
  & &  \;\;  [\attr{t'}{itemno} = \attr{t}{itemno}] \times [\attr{t'}{np} = t.np] \times\\
  & & \;  [\attr{t'}{np} > 1000] \times \rel{price}(t') \times \rel{itm}(t_2) \parallel  & \\
\end{array}
\]
\end{small}
Similarly, $Q_2(t)$ is canonized to:
\begin{small}
\[
\begin{array}{llll}
 Q_2(t) & \eq &  \parallel \sum_{t_1, t_2} [\attr{t_2}{type} = \attr{t}{type}] \times [\attr{t_2}{itemno} = \attr{t}{itemno}]  \times   & \\
  & &  \;\;  [\attr{t_1}{itemno} = \attr{t}{itemno}] \times [\attr{t_2}{itemno} = \attr{t_1}{itemno}] \times \\
  & & \; [\attr{t_1}{np} = t.np] \times [\attr{t_1}{np} > 1000] \times \rel{price}(t_1) \times \rel{itm}(t_2) \parallel  & \\
\end{array}
\]
\end{small}

At the end, since $Q_1(t)$ and $Q_2(t)$ are squashed expressions,
\setalgname is called. \setalgname finds a homomorphism from $Q_2(t)$ to $Q_1(t)$, namely
$\{t_1 \rightarrow t', t_2 \rightarrow t_2\}$ and a homomorphism from $Q_1(t)$ to $Q_2(t)$ defined
by the function $\{t' \rightarrow t_1, t_2 \rightarrow t_2\}$. Hence $Q_1$ is equivalent to $Q_2$.
To the best of our knowledge,
this is the first time that this rewrite rule is formally proved to be correct.
By modeling the semantics of SQL using \newexps, our procedure can be used to
automatically deduce the equivalence of complex rewrites.

%

\section{Evaluation}

\seclabel{evaluation}

In this section we describe our implementation and evaluation of \algname.
We first describe our implementation in~\secref{impl}. 
Then, in~\secref{rules}, we report on the two sets of query rewrite 
rules that we used in the evaluation: one set from well-known data management 
research literature, and another from a popular open-source query optimization 
framework called Calcite~\cite{calcite}. 
We summarize the evaluation results of our query equivalence checking 
algorithm in~\secref{rules} and characterize these results in~\secref{breakdown}.  
\secref{limitations} concludes with the limitations of our current implementation.

\subsection{Implementation}
\seclabel{impl}
 
We have implemented our \algname equivalence checking algorithm,
and \figref{arch} shows its architecture. 
As shown in the figure, 
Our implementation takes as input a pair of SQL queries to be checked ($Q_1$ and $Q_2$) and 
the precondition for these queries in the form of integrity constraints ($C$). 
It compiles the queries to {\newexp}s ($E_1$ and $E_2$) 
and checks the equivalence of the resulting \newexps using \algname.
We implemented \algname in \reviewermeta{Lean}, a proof assistant, 
which guarantees that if \algname returns true then the input queries are 
indeed equivalent according to our \ourring semantics.

\reviewermeta{
\algname is implemented in two components: a \newexp generator that parses the input SQL queries (expressed using the syntax shown in~\figref{sql-syntax}) to ASTs (Abstract Syntax Trees), and a converter that translates the ASTs to {\newexp}s. 
The parser is written in $440$ lines of Haskell, 
and the converter is written using $202$ lines of Lean. 
In addition, the implementation of the axioms discussed 
in \secref{axioms} and \secref{constraints} consists of $129$ lines of Lean.
\algname is implemented using $1422$ lines of Lean's metaprogramming language 
for proof search~\cite{leanmeta}. 
The parser, converter, and axiom code form the trusted code base of our implementation.
All other parts, such as the implementation of \algname, are formally verified by implementing in Lean on top of our trusted code base. }

\begin{figure}
\centering
\includegraphics[width=\columnwidth]{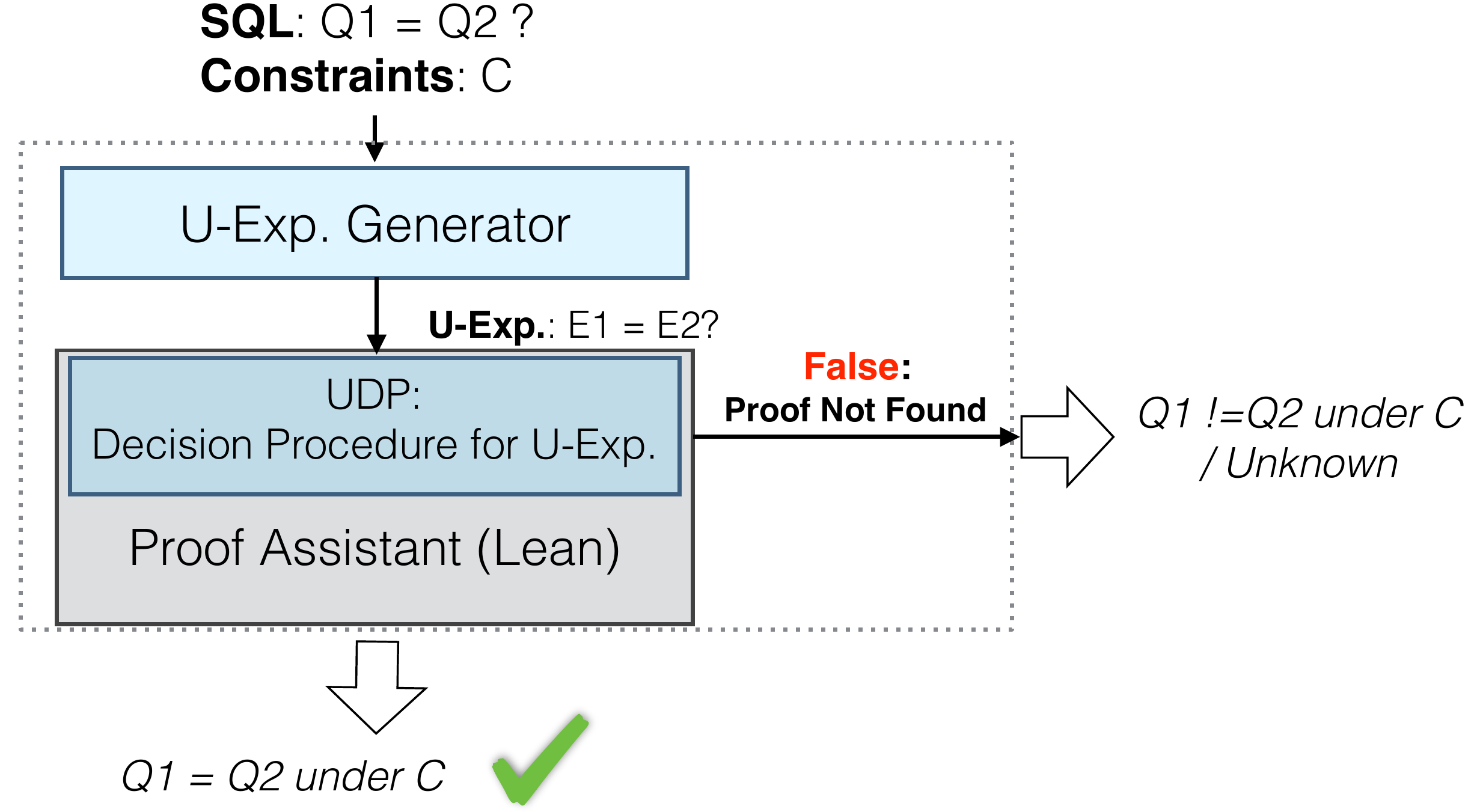}
\vspace{-8mm}
\caption{\algname implementation}
\figlabel{arch}
\vspace{-3mm}
\end{figure}

\subsection{Evaluation Summary}
\seclabel{rules}
To evaluate \algname,
we used it to prove various real-world SQL \reviewerone{queries and} rewrite rules from the following two sources \footnote{The detailed benchmark queries and rules can be found in~\cite{DBLP:journals/corr/abs-1802-02229}}:

\paragraph{Literature}. \reviewerone{We manually examined the SIGMOD papers in the last 30 years looking for rewrite rules that contain integrity constraints.}
We found $6$ rewrite rules 
from research papers published in SIGMOD 
(such as rules with mixed bag-set semantic queries from starburst~\cite{SeshadriHPLRSSS96SIGMOD}), 
technical reports, and blog articles. These rules are {\em conditional}, i.e., they all claim to be valid only
under integrity constraints that are expressible in SQL. These constraints include key constraints, foreign key constraints, and the use of indexes. 
\reviewerone{
We also include 23 rewrite rules that are interactively proven in a proof assistant from
our previous work \cite{cosette_pldi}, including the well-known magic set rewrites~\cite{SeshadriHPLRSSS96SIGMOD}. }

\paragraph{Apache Calcite}. Apache Calcite~\cite{calcite} is
an open-source query optimization framework that 
powers many data processing engines, such as Apache Hive~\cite{hive}, 
Apache Drill~\cite{drill}, and others~\cite{mapd, phoenix, flink, kylin}.
Calcite includes an extensive rule-based rewrite optimizer 
that contains $83$ rules total. 
To ensure the correctness of these rewrite rules, Calcite comes with 232 test cases, each
of which contains a SQL query, a set of input tables, and the expected results.
Passing a test case means that the Calcite rewritten query returns the correct result 
on the specific test data. 
For each Calcite's test case, we use the input query as $Q_1$ and 
the rewritten query after applying Calcite's rules
as $Q_2$.
Among these 232 pairs of SQL queries, 
39 pairs use SQL features that \algname currently supports, and 
we discuss why \algname cannot support the rest in~\secref{limitations}.
\reviewerone{Note that the Calcite test cases lead us to verify
  pairs of {\em queries} rather than rules.  For example, one of the test cases checks for
  $R \Join S = S \Join R$, where $R,S$ are concrete tables rather than arbitrary SQL expressions.
 As Calcite implements its rewrite rules in Java code rather than our input language as shown in~\figref{sql-syntax}, we instead treat every pair of queries in the test cases as query instances by replacing the concrete table names with general SQL expressions using our syntax, while retaining any integrity constraints and ignoring the actual contents of the tables. We then send the queries to our system and ask it to prove the semantic equivalence of the two queries for all possible table contents.}

Among the set of rules used in our evaluation, 8 from the literature and 
2 from Calcite require database integrity constraints as preconditions. 
Furthermore, 14 from the literature and 12 from 
Calcite were not conjunctive query rewrites.

\figref{exp_summary} summarizes our evaluation result.
Among the $6$ conditional query rewrites from the literature,
\algname can automatically prove all of them. 
Among Calcite's $39$ query rewrite rules that use SQL features 
supported by \algname, 
$33$ (i.e., $85\%$) of them are automatically proved.
$5$ unproven rewrite rules involve integer arithmetic and string casting, which \algname
currently does not support.
The last unproven rewrite rule involves two very long queries, 
and \algname does not return a result after running for 30 minutes.

\reviewerone{
\paragraph{Previously Documented Bugs.} We tried using our system to prove the
  count bug~\cite{countBug}. As expected, our system failed to
  prove equivalence within the time limit of 30 minutes; two other bugs in the literature~\cite{pgbug,
    oraclebug} are based on the NULL semantics, which we currently do
  not support. As our prior work
  work~\cite{cosette_cidr} already shows how to use a model checker to find counterexamples to invalidate such rules, our current system instead focuses on proving equivalent rules instead.}

\begin{figure}
\centering
\begin{tabular}{|l|l|l|l|l|}
\hline
\textbf{Dataset} & No. of & No. of Rules& No. of& No. of\\
                 & Rules& Supported & Proved  & Unproved   \\
                 &      &           & Rules   & Rules  \\\hline
  Literature       & \reviewerthree{29}                     & \reviewerthree{29}                             & \reviewerthree{29}                            & 0         \\ \hline
  Calcite          & 232                   & 39                              &  33                           &  6                           \\ \hline
 \reviewerone{Bugs} & \reviewerone{3} & \reviewerone{1} & \reviewerone{0} &  \reviewerone{1} \\ \hline
\end{tabular}
\caption{Summary of proved and unproved rewrite rules}
\figlabel{exp_summary}
\end{figure}
\vspace{-2mm}

\subsection{Characterizing Results}
\seclabel{breakdown}

\begin{figure}
\centering
\resizebox{\columnwidth}{!}{%
\begin{tabular}{|l|l|l|l|l|l|l|}
\hline
\multicolumn{1}{|c|}{\textbf{Dataset}} & 
\multicolumn{1}{c|}{\begin{tabular}[c]{@{}c@{}}\textbf{Proved} \\\textbf{Total} \end{tabular}} & 
\multicolumn{1}{c|}{\textbf{UCQ}} & 
\textbf{Cond.} & 
\multicolumn{1}{c|}{\begin{tabular}[c]{@{}c@{}}\textbf{Grouping,} \\ 
\textbf{Aggregate,}\\ 
\textbf{and Having}\end{tabular}} & 
\multicolumn{1}{c|}{\begin{tabular}[c]{@{}c@{}} \texttt{DISTINCT} \\  
\textbf{in} \\ \textbf{Subquery} \end{tabular}} \\

\hline
Literature & 29 & 15 & 9 & 2 & 4 \\ \hline
Calcite & 34 & 21 & 2 & 11  &  1\\ \hline
\end{tabular}%
}
\caption{Characterization of proved rewrite rules, where the categories are not mutually exclusive.}
\figlabel{breakdown}
\end{figure}

As shown in~\figref{breakdown}, we categorize the proved rewrite rules based on the SQL features that were used into the following:
\begin{itemize}[leftmargin=*, itemsep=1pt]
\item \textbf{UCQ}: Rewrites involving only unions of conjunctive queries, i.e., 
unions of \texttt{SELECT-FROM-WHERE} with conjunctive predicates. 
\item \textbf{Cond}: Rewrites that involve integrity constraints as preconditions, for instance a rewrite that is only valid in the presence of an index on a particular attribute.
\item \textbf{Grouping, Aggregate, and Having}: Rewrites that use at least one of \texttt{GROUP BY}, 
aggregate functions such as \texttt{SUM}, and \texttt{HAVING}.
\item \texttt{DISTINCT} \textbf{in Subquery}: Rewrites with \texttt{DISTINCT} in a subquery. 
\end{itemize}

As~\figref{breakdown} shows, \algname can formally prove the equivalence of 
many of the SQL rewrites described above.
The running time of \algname on all these rules are within $15$ seconds. 
Many such rewrites involve queries that are beyond UCQ, 
i.e., they are not part of the decidable fragment of SQL, 
such as the 3 rewrite rules from Starburst~\cite{PiraheshHH92}
(we described one of them in~\secref{alg-example}).
To the best of our knowledge, none of these 3 rules (along with 35 other rules that \algname proved) 
were formally proven before. 
Proving the equivalences of these rewrite rules is non-trivial: it requires reasoning equivalence of queries evaluated under 
mixed of bag and set semantics, and modeling various preconditions and subqueries that use \texttt{DISTINCT}. 

\begin{figure}
\centering
\resizebox{\columnwidth}{!}{%
\begin{tabular}{|l|l|l|l|l|l|}
\hline
\multicolumn{1}{|c|}{\textbf{Dataset}} & 
\textbf{Overall Avg.} & 
\multicolumn{1}{c|}{\textbf{UCQ}} & 
\textbf{Cond.} & 
\multicolumn{1}{c|}{\textbf{\begin{tabular}[c]{@{}c@{}}Grouping, \\ Aggregate,\\ and Having\end{tabular}}} & 
\multicolumn{1}{c|}{\begin{tabular}[c]{@{}c@{}} \texttt{DISTINCT} \\ \textbf{in} \\ \textbf{Subquery} \end{tabular}} \\ \hline
\reviewertwo{Literature} & \reviewertwo{6594.3}  & \reviewertwo{3480.8} & \reviewertwo{9983.9}  & \reviewertwo{8628.1}  &  \reviewertwo{8223.7}   \\ \hline
\reviewertwo{Calcite} & \reviewertwo{4160.4}  & \reviewertwo{2704.9} &  \reviewertwo{6429.0} &  \reviewertwo{6909.4}  &  \reviewertwo{6427.7}   \\ \hline
\end{tabular}%
}
\caption{\reviewertwo{\algname execution time (ms)}}
\figlabel{loc}
\vspace{-3mm}
\end{figure}

\reviewertwo{
\figref{loc} shows the run time of  \algname for proving the rewrites in each category. For the rewrite rules from Literature, \algname takes $6594.3$ ms on average. For the ones from Calcite, \algname takes $4160.4$ ms on average. As expected, \algname takes longer time on rewrites with rich SQL features such as integrity constraints, grouping and aggregate, and \texttt{DISTINCT} in subquery.}

\reviewerone{
An interesting question to ask is whether converting a \newexp to \firstnf increases its size significantly in practice. Theoretically, Rule~\ref{rule:distr-prod-plus-r} and Rule~\ref{rule:distr-prod-plus-l} can increase the size of \newexp exponentially. We recorded the sizes of \newexp before and after converting them to \firstnf.
\newexp sizes increase by $4.1\%$ on average in the Literature category, and increase by $0.7\%$ on average in Calcite. Despite the exponential growth at the worst case, our evaluation shows that the growth of {\newexp}s after normalization is not a big concern.
}


\paragraph{Comparison to {\oldsys}.} We also compare \algname with \oldsys~\cite{cosette_pldi}. 
\algname supports a wider range of SQL queries and 
provides more powerful \emph{automated} proof search compared with \oldsys.
In fact, \oldsys can only express \reviewermeta{$61$} out of \reviewermeta{$69$} rewrite rules that \algname proved 
(\figref{exp_summary}) as \oldsys does not support all types of database integrity constraints that \algname supports.
For the \reviewermeta{$61$} rules that \oldsys can express, only \reviewermeta{$17$} of them (Ex.~\ref{ex-index}) was manually proven by \oldsys, and none of them can be proved automatically. 
As a comparison, {\oldsys}'s manual proof script contains $320$ lines of Coq to prove Ex.~\ref{ex-index}, 
in contrast to \algname automatically proving this rewrite rule. 

\subsection{Limitations}
\seclabel{limitations}
\paragraph{Unsupported SQL Features.} 
\algname currently does not support SQL features
such as \texttt{CASE}, \texttt{UNION} (under set semantics), 
\texttt{NULL}, and \texttt{PARTITION BY}. 
The queries in the rest of Calcite dataset \reviewerthree{($193$ rules)} contains at least one of these features and 
hence cannot be processed by our current prototype.
Many of these features can be handled by syntactic rewrites. 
For example, \texttt{UNION} can be
rewritten using \texttt{UNION ALL} and \texttt{DISTINCT}. 
We believe further engineering will enable us to support the majority of 
the remaining rewrite rules and they do not represent any fundamental obstacles to our approach.

\paragraph{Unproven Rules.} 
There are several rewrites that use only the supported SQL features 
but \algname still fails to find proofs for them ($6$ out of $39$ in the Calcite dataset). 
An example from Calcite is shown below:

\begin{lstlisting}[style=sql]
SELECT *                              -- Q1
FROM (SELECT * FROM EMP AS EMP 
      WHERE EMP.DEPTNO = 10) AS t 
WHERE t.DEPTNO + 5 > t.EMPNO;

SELECT *                              -- Q2
FROM (SELECT * FROM EMP AS EMP0 
      WHERE EMP0.DEPTNO = 10) AS t1 
WHERE 15 > t1.EMPNO; 
\end{lstlisting}

Proving the above rewrite requires modeling the semantics of integer arithmetic
(which is undecidable in general), while other rules require modeling the semantics of 
string concatenation and conversion of strings to dates.
 We leave supporting such cases as future work. 



\section{Related Work}

\noindent \textbf{Containment and equivalence of relational queries.} 
Relational query containment and equivalence is a well studied topic in data management research. Equivalence of general relational queries has been proven to be undecidable~\cite{Trakhtenbrot50DANUSSR}, and subsequent research has  focused on identifying decidable fragments of SQL, such as under set~\cite{cqDecidable, SagivY80JACM} or bag semantics~\cite{CohenNS99, JayramKV06PODS, ChaudhuriV93, IoannidisR95}. 
As mentioned in~\secref{intro}, this line of work has focused on the theoretical aspects of the problem and has led to very few implementations,
most of which has been restricted to applying the chase procedure to conjunctive
queries for query optimization~\cite{benediktSurvey}. A recent work~\cite{GuagliardoL17} proposes a new semantics to model many features of SQL (including NULL semantics), despite the lack of evaluation using real-world benchmarks and a usable implementation.

\noindent \textbf{Semantic query optimization.}
Semantic query optimization is an important topic in query processing. While typical database engines optimizes queries using rule-based~\cite{PiraheshHH92} or cost-based~\cite{cascades, systemr} techniques, the line of work mentioned above has led to alternative approaches, most notably the chase and backchase (C\&B) algorithm \cite{PopaT99, PopaDST00, DeutschPT99, PennTR}, 
which guarantees to find a minimal semantic equivalent query for conjunctive queries under constraints. 
While our algorithm (Alg.~\ref{alg:tonf}) bears resemblance to C\&B, our work fundamentally differs in that our goal is the find a formal, machine-checkable proof for the equivalence of two queries using a small number of axioms, while C\&B aims to find a minimal equivalent query to the input.
Second, our algorithm is sound for general SQL queries and complete for UCQ under set and bag semantics, while the original C\&B are applicable only to CQ evaluated under set semantics~\cite{PopaT99, PopaDST00, DeutschPT99}, and the bag semantics version~\cite{PennTR} is sparse in formal details and proofs.

\noindent \textbf{Formalization of SQL semantics.}
The most related SQL formalizations include \cite{cosette_pldi, cosette_cidr, 
cosette_demo, GuagliardoL17, AuerbachHMSS17SIGMOD}.
{\oldsys}~\cite{cosette_pldi, cosette_cidr, cosette_demo} formalized
K-relation in the Coq proof assistant using univalent types
in Homotopy Type Theory (HoTT) ~\cite{hottbook}. 
Compared to {\oldsys}, {\algname} has a much smaller axiomatic foundation 
and yet more powerful decision procedures that can find proof scripts for wider range of SQL queries.
We have already discussed \cite{GuagliardoL17, AuerbachHMSS17SIGMOD} in detail in \secref{overview}.  
Related formalizations of SQL or SQL like declarative languages in SMT solvers 
include Qex \cite{VeanesTH10LAPR16, VeanesGHT09ICFEM}, a tool for verifying 
the equivalences of Spark programs \cite{GrossmanCIRS17}, Mediator, 
a tool for verifying database driven applications \cite{WangDLC18}, and Blitz, a tool for synthesize big data queries \cite{SchlaipferRLS17}. 
Unlike \algname, Qex is used for test generation. The Spark verifier~\cite{GrossmanCIRS17} can automatically verify the equivalences for a small set of Spark programs. However, it cannot be applied to SQL queries due to its syntactical restrictions. Mediator focuses on verifying transactions and programs that make updates to databases and is orthogonal to our work. Blitz~\cite{SchlaipferRLS17} can only check SQL query equivalence up to bounded size inputs and is not a full verifier.


\section{Conclusion}

In this paper we presented \ourring, a new semantics for SQL based on unbounded semirings. Using only a few axioms, \ourring can model many SQL features including integrity constraints, \reviewertwo{which is not handled in prior work. To show the usefulness of \ourring, we have developed a novel algorithm, \algname, for checking the equivalence of SQL queries and have used it to prove the validity of $62$ real-world SQL rewrites, many of which were proven for the first time. As future work, we plan to support more SQL features and other non-relational data models such as Hive and Spark.}


\bibliographystyle{abbrv}
\bibliography{paper} 


\newpage
\begin{appendix}

In this appendix we provide details on translating 
\sem to \newexp. The translation has two stages. The first stage 
converts \sem to \hottir (\secref{app-to-hottir}), an intermediate representation (IR) for \sem. 
The major difference between \sem and \hottir is the data model: 
{\sem}'s data model is named, 
thus the references to attributes uses attribute names; 
and {\hottir}'s data model is unamed, thus the references to attributes 
use path expressions (\secref{data-model}). 
The second stage converts \hottir to \newexp (\secref{app-denote}). 

\section{Data Model of \sem and \hottir}
\seclabel{data-model}

\subsection{Data Model of \sem}
We first describe how schemas for relations and tuples are modeled
in \sys. Both of these foundational concepts are from relational 
theory~\cite{Codd70CACM} that \sem builds upon. Note, the data model of 
SQL that we used here is roughly the same as \cite{cosette_pldi}, 
with the exception of replacing univalent type with \ourring .

\paragraph{Schema and Tuple}
Conceptually, a database schema is an unordered bag of attributes represented as
pairs of $(n, \tau)$, where
$n$ is the attribute name, and $\tau$ is the type of the attribute. 
We assume that each of the SQL types can be translated into 
their corresponding Lean data type.

In \sys, a user can declare a schema $\sigma$ with three attributes as follows:
\begin{small}
\[ \texttt{schema $\schema(\texttt{Name:string, Age:int, Married:bool})$;}  \]
\end{small}
A database tuple is a collection of values that conform to a
given schema. For example, the following is a tuple with the 
aforementioned schema:
\begin{small}
\[ \{\text{Name}:\text{``Bob''}; \; \text{Age}:52; \; \text{Married}: \true \} \]
\end{small}
Attributes are accessed using record syntax. For instance
 $t.\text{Name}$ returns ``Bob'' where $t$ refers to the tuple above. 

To enable users express rewrite rules over arbitrary
schemas, we distinguish between {\em concrete} schemas, in which all of 
their attributes are known, from {\em generic} schemas, which might contain
unknown attributes that are denoted using {\tt ??}. 
For example, the following rewrite rule: 
\begin{small}
\begin{align*}
\texttt{SELECT x.a as a FROM r x WHERE TRUE AND x.a = 10} &  \\
 \equiv \quad \texttt{SELECT x.a as a FROM r x WHERE x.a = 10} & 
\end{align*}
\end{small}
\noindent holds for any table with a schema containing at least the integer attribute $a$. 
In \sys, this is expressed as a generic schema that is declared as:
\begin{small}
\[ \texttt{schema $\schema(\texttt{a:int, ??})$} \]
\end{small}

\paragraph{Relation}
In \sem a relation is modeled as a function from tuples to \ourring. 

\subsection{The data model of \hottir}

\begin{figure}[t]
\begin{small}
\[
\begin{array}{llll}
  \type \in \texttt{Type}                 & ::=     & \sqlInt \sep \sqlBool \sep \sqlString \sep \; \ldots \\ 
  \denote{\sqlInt}                        & ::=     & \Integer \\
  \denote{\sqlBool}                       & ::=     & \Bool \\
  \ldots & & \\

  \schema \in \Schema                     & ::=     & \emptySchema \\
                                          & \; \mid & \leaf{\type} \\
                                          & \; \mid & \node{\schema_1}{\schema_2} \\ \\

  \Tuple \; \emptySchema                  & ::=     & \Unit \\
  \Tuple \; (\node{\schema_1}{\schema_2}) & ::=     & \Tuple \; \schema_1 \times \Tuple \; \schema_2 \\
  \Tuple \; (\leaf{\type})                & ::=     & \denote{\type} \\

\\
  \Attr \; \schema \; \type            & ::=     & \Tuple \; \schema \rightarrow \denote{\type} 
\end{array}
\]
\end{small}
\caption{\hottir's implementation of \sem{'s} Data Model}
\label{fig:data-model}
\end{figure}

\begin{figure}[t]
\begin{small}
\[
\arraycolsep=3pt
\begin{array}{lll}
\texttt{schema } \schema(\texttt{a:int, ??})\texttt{;} & \Rightarrow & 
\schema: \texttt{Schema.}  \\
  & &\texttt{a: Attribute } \schema \; \type \texttt{.} \\
\\ 
 \text{schema of }  & \Rightarrow &  \node{(\leaf{\sqlInt})}{} \\
 \texttt{SELECT "Bob" AS Name, }   & &  \qquad \; (\node{(\leaf{\sqlInt})}{} \\
 \texttt{52 AS Age, TRUE AS Married} & &  \qquad \; \qquad \; \; (\leaf{\sqlBool}))  \\
\\
\text{schema of}  & \Rightarrow &  \node{\schema_1}{\schema_2} \\
\texttt{SELECT * FROM t1 x, t2 y} & & \\
\texttt{(table t1 $\schema_1$; table t2 $\schema_2$;)}
 \end{array}
\]
\end{small}

\caption{Examples of \hottir's implementation of schemas}
\label{ex:data-model}
\end{figure}

\figref{data-model} shows {\hottir}'s implementation of 
\sem{'s} relational data model in Lean.
Schemas are modeled as a collection of types
organized in a binary tree, with each type corresponding to an attribute. 
Tuples in \sem are implemented as dependent types on a schema. As shown
in~\figref{data-model}, given a schema $s$, if $s$ is the empty schema, then 
only the empty (i.e., {\tt Unit}) tuple can be constructed from it.
Otherwise, if $s$ is a leaf node in the schema tree with type $\tau$,
then a tuple is simply a value of type $\D{\tau}$. 
Finally, if $s$ is recursively
defined using two schemas $s_1$ and $s_2$, then the resulting tuple
is an instance of a product type $\Tuple(s_1) \times \Tuple(s_2)$. An attribute
in \sem is implemented as an uninterpreted function from a tuple of its schema to
its data type.

Figure~\ref{ex:data-model} shows three examples of \hottir{'s} Lean implementation 
of schemas and tuples.
A declaration of a generic schema \texttt{$\schema$(a:int,??)} is
implemented as a Schema declaration (\texttt{$\schema$:Schema}), and an
attribute declaration (\texttt{a:Attribute $\schema$ $\type$}) in Lean. In the
second example, the output schema of the SQL query is represented as a 
tree. Lastly, the output schema of a SQL query that performs a
Cartesian product of two tables is implemented as a tree with each table's
schema as the left and right node, respectively.

\hottir implements schemas as trees.
We do so simply for engineering convenience, as
using trees allows us to easily support generic schemas in Lean.
Consider the third example in Figure~\ref{ex:data-model}, which joins
two tables with generic schema $\schema_1$ and $\schema_1$ into a table
$t$ with schema $(\node{\schema_1}{\schema_2})$. Since we know
that every tuple of $t$ has type $(\Tuple \; (\node{\schema_1}{\schema_2}))$, by the computation rule for $\Tuple$, it has type $(\Tuple \; \schema_1 \times \Tuple \; \schema_2)$.
This computational simplification enables
accesses to the components of a tuple which was generated
by joins of tables with generic schemas.
A straightforward alternative implementation is to model schemas as lists. 
Then $t$'s schema is
the list concatenation of $\schema_1$ and $\schema_2$, i.e.,
$\texttt{append}(\schema_1, \schema_2)$, and
every tuple of $t$ has type $(\Tuple \; \texttt{append}(\schema_1, \schema_2))$.
However, this cannot be further simplified computationally, because the definition
of $\mathit{append}$ gets stuck on the generic schema $\schema_1$.

\section{Translating \sem to \hottir}
\seclabel{app-to-hottir}

\figref{hottir-syntax} shows the syntax of \hottir. 
It is an unnamed version of \sem. 

\begin{figure}[h]
\centering
\[
\begin{array}{llll}
  q \in \texttt{Query} & ::=  &  t \\
        & \; \mid & \texttt{SELECT} \; p \; q      \\ 
        & \; \mid & \texttt{FROM} \; q_1, \ldots, q_n      \\
        & \; \mid & q \; \texttt{WHERE} \; b       \\
        & \; \mid & q_1 \; \texttt{UNION ALL} \; q_2   \\
        & \; \mid & q_1 \; \texttt{EXCEPT} \; q_2      \\
        & \; \mid & \texttt{DISTINCT} \; q \\
 t \in \texttt{Table} \\
 b \in \texttt{Predicate} & ::= & e_1 \; \texttt{=} \; e_2 \\
       & \; \mid &  \texttt{NOT} \; b 
            \mid b_1 \; \texttt{AND} \; b_2 
            \mid b_1 \; \texttt{OR} \; b_2 \\ 
       & \; \mid & \texttt{TRUE} \mid \texttt{FALSE} \\
       & \; \mid & \texttt{CASTPRED} \; p \; b \\
       & \; \mid & \texttt{EXISTS} \; q \\
 e \in \texttt{Expression} & ::= & \Var \; p \\
       & \; \mid & f(e_1, \ldots, e_n)   
            \mid agg(q)   \\
       & \; \mid & \texttt{CASTEXPR} \; p \; e \\
 p \in \texttt{Projection} & ::= & \Star \mid \ProjL \mid \ProjR \mid \texttt{Empty} \\
       & \; \mid &  \Compose{p_1}{p_2} \\
       & \; \mid & \Duplicate{p_1}{p_2} \\
       & \; \mid & \Evaluate{e}  
\end{array}
\]
\caption{Syntax of \hottir}
\label{fig:hottir-syntax}
\end{figure}

\figref{translating-hottsql} shows the translation rules and the scheme inference rules from \sem to \hottir. A transformation rule (\texttt{Tr$_{c}$($\ldots$)}) will translate a \sem AST to a \hottir AST given a context schema $c$. A schema inference rule will infer the schema of a query or a projection.

In \figref{translating-hottsql}, the function \texttt{topath} will convert a table alias to a path expression given a context $c$. The function \texttt{table\_schema} will look up \sem{'s} declarations and find the schema of table $t$. The function \texttt{alias\_schema} will look up the schema of a table alias given a context $c$.   

\newcommand{\hctx}{\texttt{Ctx}}

\begin{figure*}
\centering
\[
\begin{array}{lll}

\multicolumn{3}{l}{ \framebox[1.1\width]{ 
    $\trans{\hctx}{\texttt{AST}_{\sem}} \rsa \texttt{AST}_{\hottir}$ } } \\
\texttt{\color{gray}{\% Translating \sem Query}}\\
\trans{c}{t} & \rsa & t \\
 \trans{c}{\texttt{SELECT $p$ $q$}} & \rsa \quad & \texttt{SELECT $\trans{\hcons{c}{\ct{c}{q}}}{p}$ $\trans{c}{q}$}  \\
\trans{c}{\texttt{FROM $q_1$ $x_1$, $q_2$ $x_2$, $\ldots$, $q_n$ $x_n$}} & \rsa &
\texttt{FROM $\trans{c}{q_1}$, $\trans{c}{\texttt{FROM $q_2$ $x_2$, $\ldots$, $q_n$ $x_n$}}$} \\
\trans{c}{\texttt{FROM $q$ $x$}} & \rsa & \texttt{FROM } \trans{c}{q} \\
\trans{c}{\texttt{$q$ WHERE $b$}} & \rsa & \trans{c}{q} \texttt{ WHERE } \trans{\hcons{c}{\ct{c}{q}}}{b} \\
\trans{c}{\texttt{$q_1$ UNION ALL $q_2$}} & \rsa & \trans{c}{q_1} \texttt{ UNION ALL } \trans{c}{q_2} \\
\trans{c}{\texttt{$q_1$ EXCEPT $q_2$}} & \rsa & \trans{c}{q_1} \texttt{ EXCEPT } \trans{c}{q_2} \\
\trans{c}{\texttt{DISTINCT $q$}} & \rsa & \texttt{DISTINCT } \trans{c}{q} \\
\texttt{\color{gray}{\% Translating \sem Predicate}}  \\
\trans{c}{e_1 = e_2} & \rsa & \trans{c}{e_1} = \trans{c}{e_2} \\
\trans{c}{\texttt{NOT }b} & \rsa &  \texttt{NOT }\trans{c}{b} \\
\trans{c}{b_1 \texttt{ AND } b_2} & \rsa & \trans{c}{b_1} \texttt{ AND } \trans{c}{b_2} \\
\trans{c}{\texttt{TRUE}} & \rsa & \texttt{TRUE} \\
\trans{c}{\texttt{FALSE}} & \rsa & \texttt{FALSE} \\
\trans{c}{\metapred(x_1, x_2, \ldots, x_n)} & \rsa & \texttt{CASTPRED } (\lkup{c}{x_1}, \lkup{c}{x_2},\; \ldots, \; \lkup{c}{x_n}) \; \metapred\\
\trans{c}{\texttt{EXISTS }q} & \rsa & \texttt{EXISTS }\trans{c}{q}\\
\texttt{\color{gray}{\% Translating \sem Expression}} \\
\trans{c}{x.a} & \rsa & \texttt{P2E $\lkup{c}{x}$.$a$ } \\
\trans{c}{f(e_1, \ldots, e_n)} & \rsa & f(\trans{c}{e_1}, \ldots, \trans{c}{e_n}) \\
\trans{c}{\agg(q)} & \rsa & \agg(\trans{c}{q}) \\
\texttt{\color{gray}{\% Translating \sem Projection}}\\
\trans{c}{\texttt{*}} & \rsa & \texttt{*} \\
\trans{c}{\texttt{$x$.*}} & \rsa & \texttt{$\lkup{c}{x}$.*} \\
\trans{c}{p_1, \; p_2} & \rsa &\trans{c}{p_1}, \; \trans{c}{p_2}  \\
\\
\multicolumn{3}{l}{ \framebox[1.1\width]{ 
    $\ct{\hctx}{\texttt{AST}_{\sem}} \rsa \hctx$ } } \\
\ct{c}{t} & \rsa & \sof{t} \\
\ct{c}{\texttt{SELECT $p$ $q$}} & \rsa &  \ct{\hcons{c}{\ct{c}{q}}}{p} \\
\ct{c}{\texttt{FROM $q_1$ $x_1$, $q_2$ $x_2$, $\ldots$, $q_n$ $x_n$}} & \rsa &
\hcons{\ct{c}{q_1}}{\ct{c}{q_2, \ldots, q_n}} \\
\ct{c}{\texttt{FROM $q$ $x$}} & \rsa & \ct{c}{q} \\
\ct{c}{\texttt{$q$ WHERE $b$}} & \rsa & \ct{c}{q} \\
\ct{c}{\texttt{$q_1$ UNION ALL $q_2$}} & \rsa & \ct{c}{q_1} \\
\ct{c}{\texttt{$q_1$ EXCEPT $q_2$}} & \rsa & \ct{c}{q_1} \\
\ct{c}{\texttt{DISTINCT $q$}} & \rsa & \ct{c}{q} \\
\ct{c}{\texttt{*}} & \rsa & c \\
\ct{c}{\texttt{$x$.*}} & \rsa & \sofc{c}{x} \\
\ct{c}{e \texttt{ AS } a} & \rsa & \typeof{c}{e} \\
\ct{c}{p_1, \; p_2} & \rsa & \hcons{\ct{c}{p_1}}{\ct{c}{p_2}} 
\end{array}
\]
Note: The Schema Inference Rule of \texttt{FROM $\ldots$} also add the table alias into context for future look up, we omit that for brevity.
\caption{Transformation Rules and Schema Inference Rules for Translating \sem to \hottir}
\figlabel{translating-hottsql}
\end{figure*}

\section{Translating \hottir{} to \outputLang}
\seclabel{app-denote}

\figref{full-denote-query} shows the denotational semantics for all constructs
in \hottir.

\begin{figure*}[t]
\centering
\[
\begin{array}{llll}

   \multicolumn{3}{l}{ \framebox[1.1\width]{ 
    $\denoteQuery{\Context}{\query}{\schema} : \Tuple \; \Context \rightarrow \Tuple \; \schema \rightarrow \Type $} }   & \texttt{(* $\Query$ *)} \\ \\

  \denoteQuery{\Context}{table}{\schema} & \teq &
    \intros \denoteTable{table} \; t \\

  \denoteQuery{\Context}{\SELECT{\proj}{\query}}{\schema} & \teq & 
    \multicolumn{2}{l}{
    \intros \sum_{\tuple':\Tuple \; \schema'}{
      (\denoteProj{\proj}{\node{\Context}{\schema'}}{\schema} \; \mkPair{\context}{\tuple'} = \tuple) \times
      \denoteQuery{\Context}{\query}{\schema'}} \; \context \; \tuple'} \\
  
  \denoteQuery{\Context}{\FROM{\query_1, \query_2}}{\node{\schema_1}{\schema_2}} & \teq &
    \intros
      \denoteQuery{\Context}{\query_1}{\schema_1} \; \context \; \fst{\tuple} \times
      \denoteQuery{\Context}{\query_2}{\schema_2} \; \context \; \snd{\tuple} \\

  \denoteQuery{\Context}{\FROM{\query}}{\schema} & \teq & 
    \intros \callQuery{\Context}{\query}{\schema} \\

  \denoteQuery{\Context}{\WHERE{\query}{\pred}}{\schema} & \teq & 
    \intros
      \callQuery{\Context}{\query}{\schema} \times
      \denotePred{\node{\Context}{\schema}}{\pred} \; \mkPair{\context}{\tuple} \\

  \denoteQuery{\Context}{\UNIONALL{\query_1}{\query_2}}{\schema} & \teq & 
    \intros
      \callQuery{\Context}{\query_1}{\schema} +
      \callQuery{\Context}{\query_2}{\schema} \\

  \denoteQuery{\Context}{\EXCEPT{\query_1}{\query_2}}{\schema} & \teq & 
    \intros
      \callQuery{\Context}{\query_1}{\schema} \times
      (\unot{(\callQuery{\Context}{\query_2}{\schema})}) \\

  \denoteQuery{\Context}{\DISTINCT{\query}}{\schema} & \teq & 
    \intros \merely{\callQuery{\Context}{\query}{\schema}} \\ \\

   \multicolumn{3}{l}{ \framebox[1.1\width]{$ \denotePred{\Context}{\pred} 
   :  \Tuple \; \Context \rightarrow \Type $  } } & \texttt{(* $\Pred$ *)}\\
  \\
  \denotePred{\Context}{e_1 = e_2} & \teq & 
    \intro(\callExpr{\Context}{e_1}{\type} = \callExpr{\Context}{e_2}{\type}) \\

  \denotePred{\Context}{\AND{\pred_1}{\pred_2}} & \teq & 
    \intro \callPred{\Context}{\pred_1} \times \callPred{\Context}{\pred_2} \\

  \denotePred{\Context}{\OR{\pred_1}{\pred_2}} & \teq & 
    \intro \merely{\callPred{\Context}{\pred_1} + \callPred{\Context}{\pred_2}} \\

  \denotePred{\Context}{\NOT{\pred}} & \teq & 
    \intro \unot{(\callPred{\Context}{\pred})}   \\

  \denotePred{\Context}{\EXISTS{\query}} & \teq & 
    \intro \merely{\sum_{t:\Tuple \; \schema}{\callQuery{\Context}{\query}{\schema}}} \\

  \denotePred{\Context}{\FALSE} & \teq & \intro \zero \\

  \denotePred{\Context}{\TRUE} & \teq & \intro \one \\
 
  \denotePred{\Context}{\CastPred{\proj}{\pred}}  & \teq  &
    \intro \denotePred{\Context'}{\pred} \; (\denoteProj{\proj}{\Context}{\Context'} \; \context) \\

   \\
  \multicolumn{3}{l}{ \framebox[1.1\width]{$ \denoteExpr{\Context}{\expr}{\type} :
    \Tuple \; \Context \rightarrow \denote{\type} $}  } & \texttt{(* $\Expr$ *)}  \\ \\

  \denoteExpr{\Context}{\Var \; \proj}{\type} & \teq & 
     \intro \denoteProj{\proj}{\Context}{\leaf \; \type} \; \context   \\

  \denoteExpr{\Context}{f(e_1, \ldots)}{\type} & \teq & 
    \intro  \denote{f}(\callExpr{\Context}{e_1}{\type_1}, \ldots )  \\

  \denoteExpr{\Context}{agg(\query)}{\type'} & \teq & 
    \intro \denote{agg} \; ( \denoteQuery{\Context}{\query}{\leaf \; \type} \; g ) \\

  \denoteExpr{\Context}{\CastExpr{\proj}{\expr}}{\type} & \teq &
    \intro \denoteExpr{\Context'}{\expr}{\type} \; (\denoteProj{c}{\Context}{\Context'} \; g) \\ \\

  \multicolumn{3}{l}{ \framebox[1.1\width]{$ 
    \denoteProj{\proj}{\Context}{\Context'} : \Tuple \; \Context \rightarrow \Tuple \; \Context'$ } }  & \texttt{(* $\Proj$ *)}  \\ \\
 
  \denoteProj{*}{\Context}{\Context} & \teq  &  \intro \context  \\

  \denoteProj{\ProjL}{\node{\Context_0}{\Context_1}}{\Context_0} & \teq & \intro \fst{\context} \\

  \denoteProj{\ProjR}{\node{\Context_0}{\Context_1}}{\Context_1} & \teq & \intro \snd{\context} \\

  \denoteProj{\Empty}{\Context}{\emptySchema} & \teq & \intro \zero \\

  \denoteProj{\Compose{\proj_1}{\proj_2}}{\Context}{\Context''} & \teq & \intro
    \denoteProj{\proj_2}{\Context'}{\Context''}\;(\denoteProj{\proj_1}{\Context}{\Context'} \; \context) \\

  \denoteProj{\Duplicate{\proj_1}{\proj_2}}{\Context}{\node{\Context_0}{\Context_1}} & \teq & 
    \intro (\denoteProj{\proj_1}{\Context}{\Context_0} \; \context, \; \denoteProj{\proj_2}{\Context}{\Context_1} \; \context) \\

  \denoteProj{\Evaluate{\expr}}{\Context}{\leaf \; \type} & \teq & \intro \callExpr{\Context}{\expr}{\type} \\

\end{array}
\]
\caption{Denotational Semantics of \hottir}
\label{fig:full-denote-query}
\end{figure*}

\end{appendix}


\end{document}